\documentclass[12pt]{iopart}

\usepackage{amssymb}
\usepackage{CJKutf8}
\usepackage{graphicx}
\usepackage[numbers,sort&compress]{natbib}
\usepackage{subcaption}
\usepackage{ulem}
\usepackage{xcolor}

\begin{document}
\begin{CJK}{UTF8}{min}

\title[Encoder--decoder-predicted evolution of a splashing drop]{Prediction of the morphological evolution of a splashing drop using an encoder--decoder}

\author{Jingzu Yee$^1$, Daichi Igarashi (五十嵐大地)$^1$, Shun Miyatake (宮武駿)$^1$, and Yoshiyuki Tagawa (田川義之)$^{1,2,*}$}

\address{$^1$ Department of Mechanical Systems Engineering, Tokyo University of Agriculture and Technology, Koganei, Tokyo 184-8588, Japan}
\address{$^2$ Institute of Global Innovation Research, Tokyo University of Agriculture and Technology, Koganei, Tokyo 184-8588, Japan}
\address{$^*$ Author to whom any correspondence should be addressed.}
\ead{tagawayo@cc.tuat.ac.jp}
\end{CJK}

\begin{abstract}
The impact of a drop on a solid surface is an important phenomenon that has various implications and applications.
However, the multiphase nature of this phenomenon causes complications in the prediction of its morphological evolution, especially when the drop splashes.
While most machine-learning-based drop-impact studies have centred around physical parameters, this study used a computer-vision strategy by training an encoder--decoder to predict the drop morphologies using image data.
Herein, we show that this trained encoder--decoder is able to successfully generate videos that show the morphologies of splashing and non-splashing drops.
Remarkably, in each frame of these generated videos, the spreading diameter of the drop was found to be in good agreement with that of the actual videos.
Moreover, there was also a high accuracy in splashing/non-splashing prediction.
These findings demonstrate the ability of the trained encoder--decoder to generate videos that can accurately represent the drop morphologies. This approach provides a faster and cheaper alternative to experimental and numerical studies.
\end{abstract}

\vspace{2pc}

\section{Introduction}
\label{sec:intro}

``Constant dripping wears through the stone'' is a famous Chinese idiom that means ``perseverance yields success''.
This idiom perfectly describes the phenomenon of the impact of a drop on a solid surface, which might appear to be insignificant but actually has great implications and many applications in both nature and industry \cite{hu2022morphology, sun2022stress, modak2020drop, breitenbach2018drop, de2015wettability}.
This is especially true when splashing occurs, i.e. when the impacting drop breaks up and ejects secondary droplets \cite{riboux2017boundary, riboux2014experiments, yokoyama2022droplet, hatakenaka2019magic, rioboo2001outcomes} instead of just spreading over the surface until it reaches a maximum diameter \cite{liu2019maximum, gordillo2019theory, lin2018impact, huang2018energetic, clanet2004maximal}.
To name just a few of the implications of splashing, in nature, it is the main cause of erosion and the propagation of contaminants, while in industry, it can cause visible decreases in printing and paint quality \cite{fernandez2017splash, waite2015grapevine, lohse2022fundamental}.

To minimize the adverse effects of this fascinating and beautiful phenomenon, many studies have been performed to understand the mechanisms and model the morphology of a splashing drop using first-principles approaches, including theoretical and numerical analyses \cite{markt2021high, philippi2016drop, lagubeau2012spreading, eggers2010drop}.
Because of the multiphase nature of a drop impact, the occurrence of splashing is heavily influenced by the physical parameters involved. These can be categorized into the impact conditions and the respective physical properties of the liquid drop, the solid surface, and the ambient air \cite{vo2021mediation, liu2021role, de2018self, gordillo2018dynamics, josserand2016drop, yarin2006drop}.
Most often, the relationships between the occurrence of splashing and each of these parameters are not simply monotonic, but rather there is an interplay between two or more of them.
Sometimes, a given parameter can either promote or suppress splashing depending on other parameters \cite{usawa2021large, zhang2022surface, zhang2021reversed}.
Furthermore, splashing can happen in different ways, including as a prompt splash, corona splash, receding breakup, and magic carpet breakup \cite{yokoyama2022droplet, hatakenaka2019magic, rioboo2001outcomes}.
For these reasons, the morphological study of the splashing phenomenon can present complications.

In recent years, with the increasing availability and accessibility of data, data-driven approaches -- and machine learning in particular -- have attracted increasing attention among fluid researchers as a faster and cheaper alternative or complement to experimental and numerical studies \cite{jalali2022physics, touranakou2022particle, vlachas2022multiscale, thavarajah2021fast, fukami2021global, novati2021automating, brunton2020machine}.
Regarding drop impacts, several machine-learning-based studies have been carried out \cite{yancheshme2022dynamic, tembely2022machine, yoon2022maximum, pierzyna2021data}.
Notably, a number of studies on predicting the maximum spreading factor of a non-splashing drop under various conditions were published in 2022.
For example, Yancheshme \textit{et al.} used the random forest to predict the maximum spreading of a drop on hydrophobic and superhydrophobic surfaces \cite{yancheshme2022dynamic}.
Also, Tembely \textit{et al.} compared the performances of the linear regression model, decision tree, random forest, and gradient boost regression model on predicting the maximum spreading of a drop on surfaces of various wettabilities \cite{tembely2022machine}.
Even for droplet-particle collisions, Yoon \textit{et al.} used a multi-layer feedforward neural network (FNN) to predict the maximum spreading diameter under a significantly wide range of impact conditions \cite{yoon2022maximum}.
Other than the maximum spreading diameter, such physical parameters-based machine learning was also applied for the investigations of the drop impact force and the splashing mechanisms.
Dickerson \textit{et al.} used a physical parameters-based ensemble learning algorithm to predict the drop impact force on concave targets \cite{dickerson2022predictive}.
As for splashing mechanisms, Pierzyna \textit{et al.} \cite{pierzyna2021data} successfully improved the splashing threshold proposed by Riboux and Gordillo \cite{riboux2014experiments}.
Although the machine learning models in these studies showed excellent performances, they were designed to use physical parameters as inputs and outputs.

As the next step, this study aims to predict the evolving dynamics of the morphology of a splashing drop during the impact.
For that, we focused on the application of image data to explore the possibility of using computer vision -- i.e. the ability of machine learning to process and generate images -- as a strategy for tackling the complications of predicting the morphology of a splashing drop.
This was inspired by Yee \textit{et al.}, who attempted to understand the splashing mechanism through the extraction of morphological features in splashing drops by visualizing how a computer vision-based FNN classified splashing and non-splashing drops\cite{yee2022image}.
In this study, the machine-learning algorithm of an encoder--decoder \cite{kingma2014auto, lu2021learning, krenn2020self, kaming2021unsupervised, badrinarayanan2017segnet, khan2023encoder, khan2020toward} was trained to generate videos in the form of image sequences that can accurately represent the actual morphological evolutions of splashing and non-splashing drops during their impact on a solid surface under specific physical parameters.
During the training of the encoder--decoder, interpolation prediction is performed by withholding intermediate data points.
The purpose of this is to build an encoder--decoder that can generate accurate post-impact image sequences for the intermediate data points. This will provide a faster and cheaper alternative to experiments or numerical simulations on new data points.

The remainder of this paper is structured as follows.
The dataset used in this study are described in detail in section~\ref{sec:data}
and the detailed implementations of the encoder--decoder model in section~\ref{sec:enc_dec}.
The results and discussion are presented in section~\ref{sec:results}:
the post-impact image sequences generated by the trained encoder--decoder are qualitatively evaluated in section~\ref{sec:gen_img_seq}, and they are quantitatively evaluated based on the spreading diameter in section~\ref{sec:spr_rad} and the accuracy of splashing/non-splashing prediction in section~\ref{sec:classification}.
Analysis on the prediction process of the trained encoder-decoder is explained in section~\ref{sec:pred_process}.
A conclusion summarizing the results is presented in section~\ref{sec:conclusion}.

\section{Dataset}
\label{sec:data}

\subsection{Data collection}
\label{sec:data_col}
\begin{figure}[tb]
\centering
\includegraphics[width=0.8\textwidth]{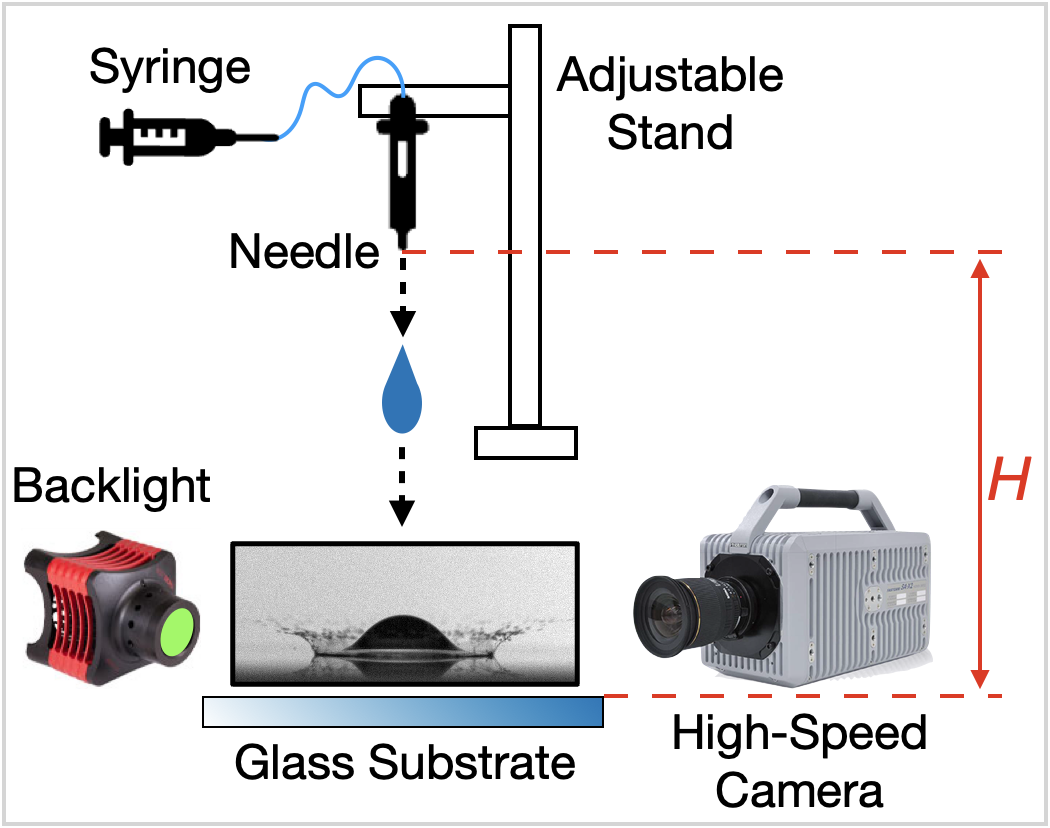}
\caption{
Experiment setup for the collection of the image data of splashing and non-splashing drops.}
\label{fig:setup}
\end{figure}
To collect the image data, experiments were performed using the setup shown in figure~\ref{fig:setup}.
A syringe was used to supply liquid ethanol (Hayashi Pure Chemical Ind., Ltd.; density $\rho = 789~\rm{kg/m^3}$, surface tension $\gamma = 2.2 \times 10^{-2}$~N/m, and dynamic viscosity $\mu = 10^{-3}$~Pa$\cdot$s) to form a drop of area-equivalent diameter $D_0 = (2.59 \pm 0.08) \times 10^{-3}$~m at a needle (internal diameter $0.97 \times 10^{-3}$~m), where it detached and impacted the hydrophilic surface of a glass substrate (Muto Pure Chemicals Co., Ltd., star frost slide glass 511611); the free-fall distance of the drop $H$ ($0.04$--$0.60$~m) was the only manipulated variable.
The measured impact velocity $U_0$ was in the range $0.82$--$3.18$~m/s, and the resulting Weber number $We$, which was computed using
\begin{equation}
\mathrm{We} = \rho U_{0}^2 D_{0}/\gamma
\label{eq:We},
\end{equation}
ranged from 63--947.
For each drop impact, a video was captured using a high-speed camera (Photron, FASTCAM SA-X) at a frame rate of 45,000~$\mathrm{s}^{-1}$ and a spatial resolution of $(1.46 \pm 0.02 )\times 10^{-5}$~m/pixel.
After a manual frame-by-frame inspection for the presence of secondary droplets, each of the videos was labelled according to the outcome: splashing or non-splashing.
A total of 249 videos were recorded: 141 splashing and 108 non-splashing.

\begin{figure}[tb]
\centering
\begin{subfigure}{0.79\textwidth}
\includegraphics[width=\textwidth]{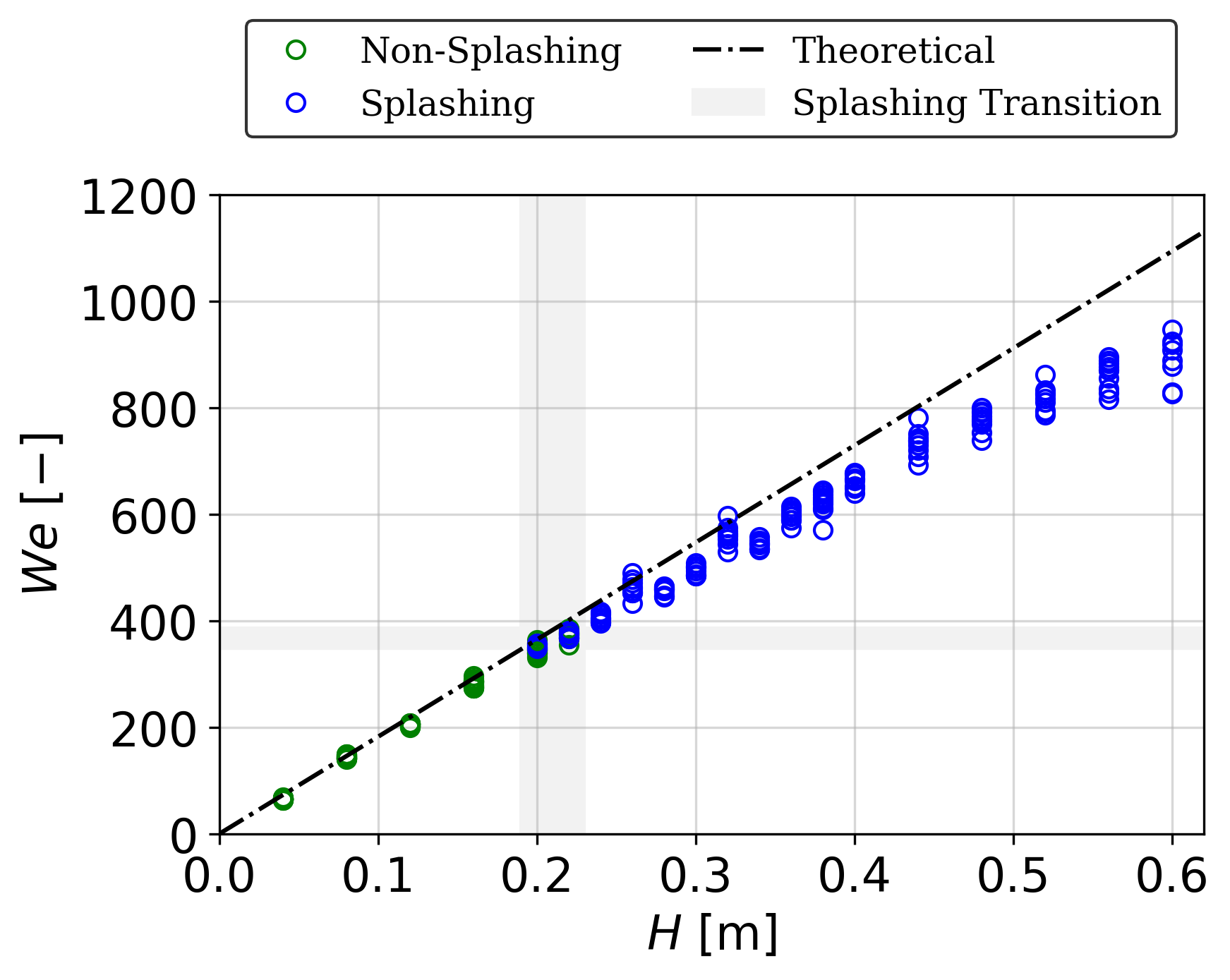}
\caption{}\label{fig:We_H}
\end{subfigure}
\\
\centering
\begin{subfigure}{0.79\textwidth}
\includegraphics[width=\textwidth]{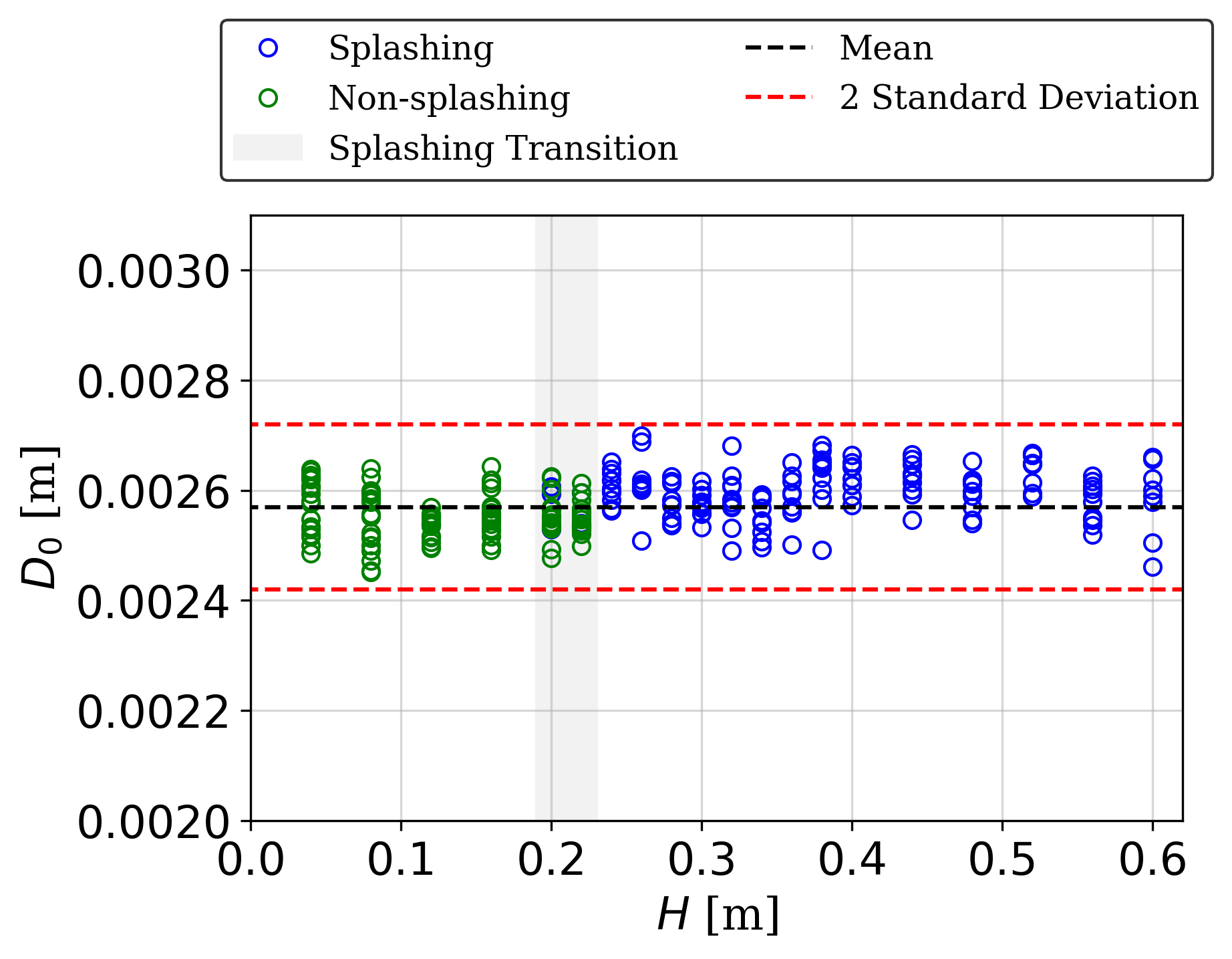}
\caption{}\label{fig:D0_H}
\end{subfigure}
\caption{
(\protect\subref*{fig:We_H}) Weber number $We$ and (\protect\subref*{fig:D0_H}) area-equivalent diameter $D_0$ versus the impact height $H$ of the collected data.}
\label{fig:D0_We_H}
\end{figure}
To evaluate the collected data in terms of the $We$ and the $D_0$, figure~\ref{fig:D0_We_H} was plotted.
Figure~\ref{fig:We_H} shows the plot of $We$ versus $H$ of the collected data.
The blue and green circles show the splashing and non-splashing data, respectively.
As shown by the areas filled with grey, the splashing thresholds in terms of impact height and Weber number were $H = 0.20$~m and $We = 348$, respectively.
Note that some impacting drops with $H$ or $We$ values greater than or equal to the splashing thresholds did not splash.
The highest values of $H$ and $We$ for a non-splashing drop were $H = 0.22$~m and $We = 386$.
Hence, there is a splashing transition within $0.20~\mathrm{m} \leq H \leq 0.22$~m or $348 \leq We \leq 386$.
The black dashed-dotted line shows the plot of the theoretical Weber number $We_{\mathrm{theo}}$ against $H$.
The equation used to compute $We_{\mathrm{theo}}$ is
\begin{equation}
\mathrm{We} = \rho U_{\mathrm{theo}}^2 D_{0}/\gamma = 2 \rho D_{0} g H/\gamma
\label{eq:We_theo},
\end{equation}
where $U_{\mathrm{theo}}$ ($=\sqrt{2gH}$) is the theoretical free-falling velocity and the gravitational acceleration $g = 9.81$~ m/s$^2$.
The collected data shows the same linear trend as the theory shown by the black dashed-dotted line.
Note that the deviation from the theory increases with $H$ because the drag increased with $H$ causing $U_0$ to deviate from $U_{\mathrm{theo}}$.

As for evaluation on the $D_0$, the plot versus $H$ of the collected data is shown in figure~\ref{fig:D0_H}.
The black and red dashed lines show the mean value and the range of two times the standard deviation, respectively.
As shown in the figure, $D_0$ of all splashing and non-splashing data lie between the range of two times the standard deviation, indicating that there was no significant bias in terms of the drop size.


\subsection{Digital image processing}
\label{sec:data_process}

Digital image processing was performed using an in-house MATLAB code to detect the drop and the glass substrate in each frame of the recorded high-speed videos.
The main image processing toolbox used for the detection is region boundary tracing \cite{gonzalez2004digital}.
The detection was enhanced using other toolboxes, such as binarization \cite{otsu1979threshold}, non-local means filtering \cite{buades2005non}, circle detection \cite{atherton1999size, yuen1990comparative}, and edge detection \cite{canny1986computational}.
Detection of the drop and the substrate was necessary for measuring the area-equivalent diameter $D_0$ and the impact velocity $U_0$ of each data.
More importantly, it also prepared the pre-impact and post-impact image sequences for the input and the output, respectively, of the encoder-decoder model.

\begin{figure}[tb]
\centering
\includegraphics[width=0.8\textwidth]{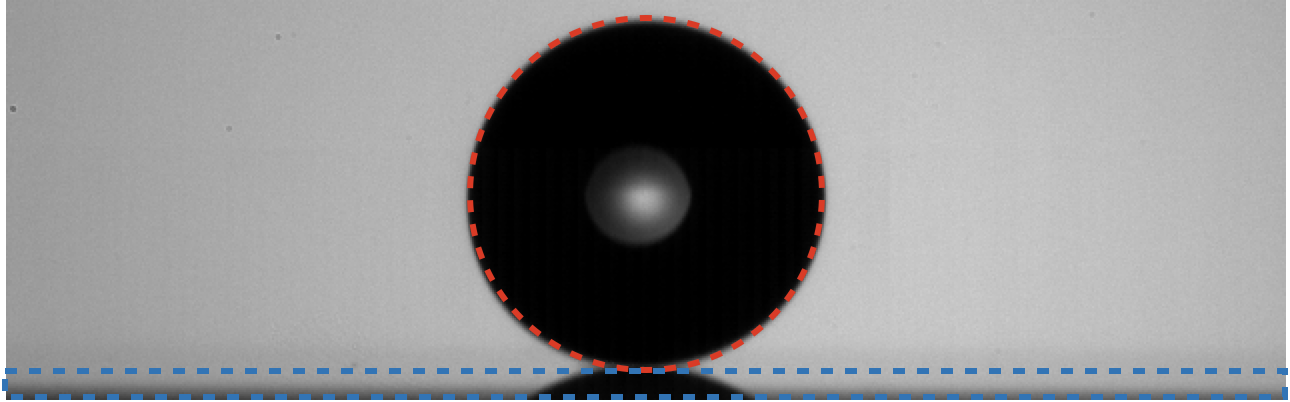}
\caption{
Illustration of the image processing performed on the collected data to identify the frame when the drop first touched the substrate ($t = 0$). 
The red dashed and the blue dashed lines show the area bound by the drop and the glass substrate, respectively, as detected by the in-house MATLAB code through region boundary tracing.
This frame was identified as $t = 0$ because the lowest point of the red dashed line is equal to the highest point of the blue dashed line.}
\label{fig:t_0}
\end{figure}
The computation of $D_0$ and $U_0$ was performed as follows.
First, the frame when the drop first touched the substrate ($t = 0$) is identified.
In the code, it was defined as the frame when the lowest point of the drop is lower than or equal to the highest point of the glass substrate.
The illustration is shown in figure~\ref{fig:t_0}.
After the frame $t = 0$ was identified, the area bounded by the drop was detected by using the toolbox for region boundary tracing to compute $D_0$ and the centre of mass.
The same procedure was applied to the previous nine frames to compute the average $D_0$ and $U_0$, which were used to compute $We$.
The related results are shown and discussed in section~\ref{sec:data_col}.

\begin{figure}[tb]
\centering
\includegraphics[width=\textwidth]{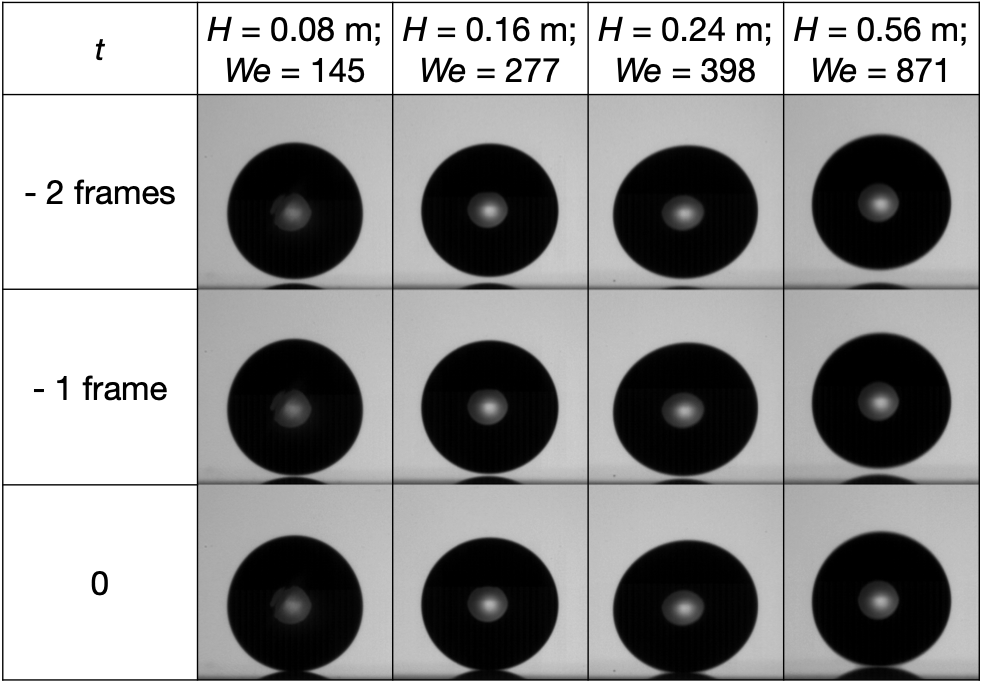}
\caption{
Examples of the input image sequences.}
\label{fig:input_img_seq}
\end{figure}
An input image sequence should indirectly provide the physical parameters of the drop impact to the encoder--decoder for the prediction of the post-impact morphology of a drop.
Therefore, an input image sequence was prepared as follows.
First, similar to the computation of $D_0$ and $U_0$, the frame when the drop first touched the substrate ($t = 0$) was identified.
The frame was then cropped from $288 \times 1024$~pixel$^2$ to $250 \times 250$~pixel$^2$, with the drop at the centre of the cropped image while the surface of the glass substrate at the bottom of the cropped image.
This was to ensure the high similarity of the frames extracted for all the collected data. 
The same procedure was applied to the two preceding frames.
Finally, the cropped frames of $t = 0$ and the two preceding frames were combined to form an input image sequence.
Some examples of the input image sequences are shown in figure~\ref{fig:input_img_seq}.
Note that all the videos were recorded at the same frame rate, thus the positions of the drop in the first two frames of a pre-impact image sequence change according to $We$, where a drop with a higher $We$ had higher positions in the first two frames due to the higher $U_0$.
This indirectly provided the encoder--decoder the information about the $We$ of a drop.

\begin{figure}[tb]
\centering
\includegraphics[width=0.5\textwidth]{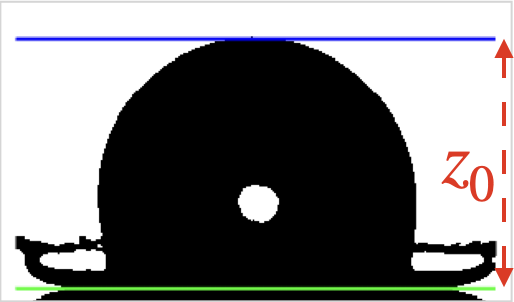}
\caption{
Illustration of the image processing performed on the collected data to compute the central height of the drop $z_0$.
The blue and green lines show the top and the surface of the glass substrate detected by the in-house MATLAB code.}
\label{fig:z_0}
\end{figure}
\begin{figure}[tb]
\centering
\includegraphics[width=\textwidth]{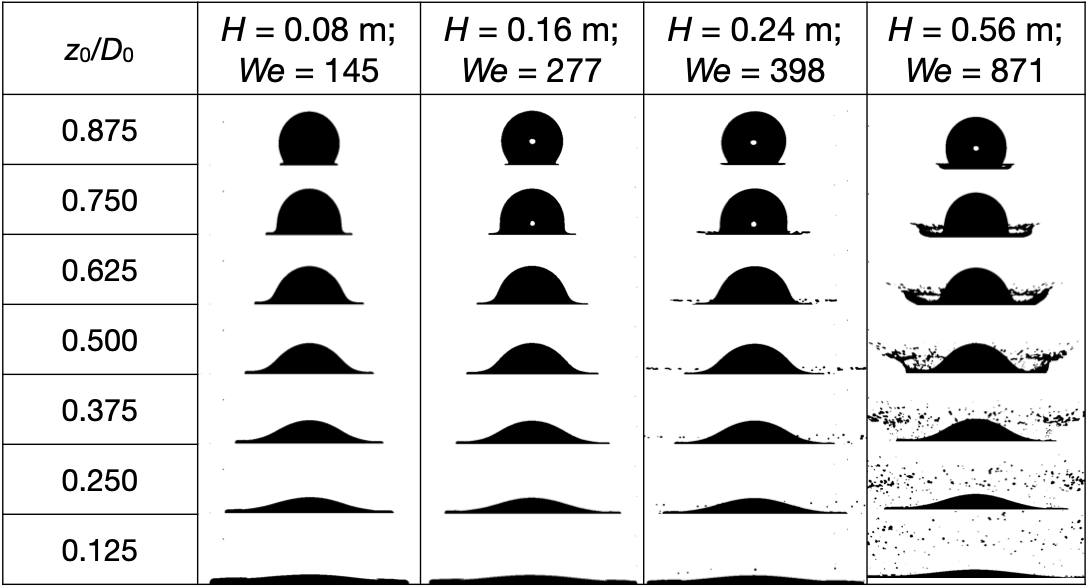}
\caption{
Examples of the output image sequences.}
\label{fig:output_img_seq}
\end{figure}

As for the output image sequence, it should show the morphology of the drop throughout the impact.
First, the top and the bottom of the drop during the impact were identified to compute the central height $z_0$.
The illustration is shown in figure~\ref{fig:z_0}.
The frames when the normalized central height of the drop $z_0/D_0 = 0.875$, $0.750$, $0.625$, $0.500$, $0.375$, $0.250$, and $0.125$ were identified and extracted.
Since $D_0$ and the spatial resolutions were kept, $z_0$ in each of these frames had the same value for all collected data.
According to Lagubeau \textit{et al.} \cite{lagubeau2012spreading}, these values of $z_0/D_0$ correspond to the pressure impact and self-similar inertial regimes.
Each of these seven frames was then cropped from $288 \times 1024$~pixel$^2$ to $640 \times 200$~pixel$^2$, with the drop at the centre of the cropped image while the surface of the glass substrate at the bottom of the cropped image.
Similar to the input image sequences, this was to ensure the high similarity of the frames extracted for all the collected data. 
Finally, these seven cropped frames were binarized and combined to form an output image sequence.
Some examples of the output image sequences are shown in figure~\ref{fig:output_img_seq}.
Note that binarization was performed so that each pixel occupied by the drop morphology had an intensity value of 0 while each pixel not occupied had an intensity value of 1.
Thus, the image-generation process can be considered as a pixel-by-pixel prediction of whether a pixel is occupied by the drop morphology.
Such a strategy was inspired by Volume of Fluid (VOF) method, a commonly used free-surface modelling technique \cite{hirt1981volume}.

\subsection{Data segmentation}
\label{sec:data_segment}

A well-trained encoder--decoder that can generate accurate post-impact image sequences for the intermediate data points can provide a faster and cheaper alternative to experiments or numerical simulations on new data points.
To examine the ability of the trained encoder--decoder to perform interpolation prediction, all image sequences of drops with $H = 0.08$, $0.16$, $0.24$, $0.32$, $0.40$, $0.48$, and $0.56$~m were excluded from the training and reserved only for the testing.
For the remaining image sequences with $H = 0.04$, $0.12$, $0.20$, $0.22$, $0.26$, $0.28$, $0.30$, $0.34$, $0.36$, $0.38$, $0.44$, $0.52$, and $0.60$~m, 90\%--10\% segmentation was performed according to each $H$ value.
Thus, 146 sets (about 59\%) and 103 sets (about 41\%) were used for training and testing, respectively.
Hereafter, the reserved $H$ and the non-reserved $H$ values are referred to as the interpolation impact height $H$ and training impact height $H$, respectively.

\section{Encoder--Decoder}
\label{sec:enc_dec}

An encoder--decoder is a type of artificial neural network that can encode higher-dimensional data into lower-dimensional representation and subsequently decode the encoded representation into the desired output of higher dimension \cite{kingma2014auto}.
It is widely used for dimensionality reduction and data generation both in fundamental research and practical applications.
An example of its applications for fundamental research is the study by Lu \textit{et al.} that designed DeepONet based on an encoder--decoder to learn diverse linear/nonlinear explicit and implicit operators \cite{lu2021learning}.
For the discovery of novel materials, Krenn \textit{et al.} used a variational autoencoder to reconstruct the molecules \cite{krenn2020self}.
Also, K\"aming \textit{et al.} applied unsupervised machine learning using an autoencoder to predict the phase transitions from the experimental data of a Haldane-like model realised with ultracold atoms \cite{kaming2021unsupervised}.
On the other hand, for practical applications, Badrinarayanan designed SegNet based on a convolutional encoder--decoder for image segmentation \cite{badrinarayanan2017segnet}.
In addition, Khan \textit{et al.} effectively utilized the framework of an encoder-decoder to extract building footprints from aerial images \cite{khan2023encoder} and to predict COVID-19 hotspots \cite{khan2020toward}.
Especially, for the prediction of COVID-19 hotspots, they successfully incorporated a hybrid model of convolutional neural network (CNN) and long short-term memory (LSTM) to process time-series data.

\subsection{Architecture}
\label{sec:architecture}

\begin{figure}[tb]
\centering
\includegraphics[width=\textwidth]{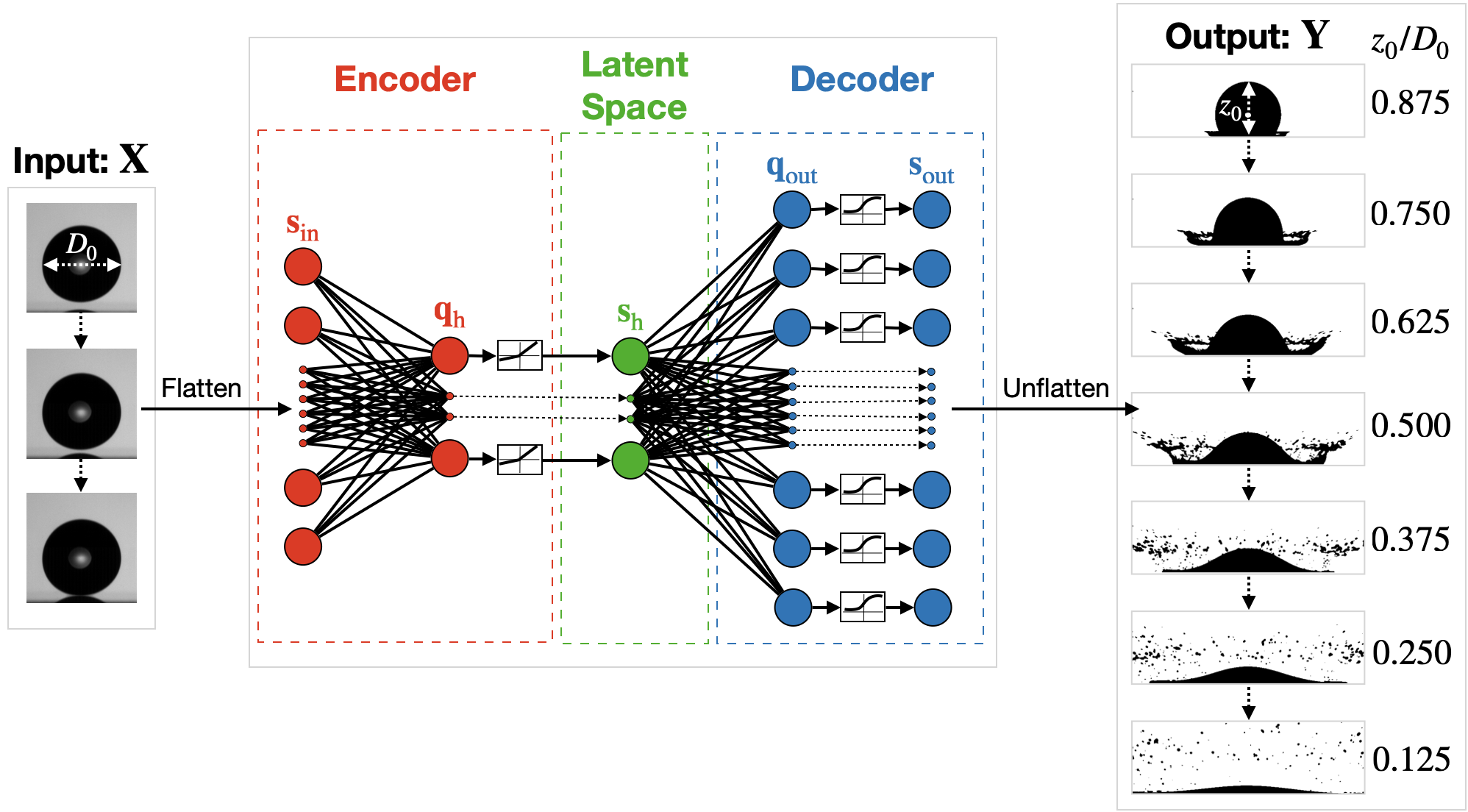}
\caption{
Architecture of the encoder--decoder trained to take a pre-impact image sequence and generate a corresponding post-impact image sequence to predict the morphological evolution of a splashing or non-splashing drop.}
\label{fig:enc_dec}
\end{figure}

In this study, an encoder--decoder with the optimized architecture shown in figure~\ref{fig:enc_dec} was trained to generate a post-impact image sequence as the output, from a pre-impact image sequence as the input.
This was implemented in the Python programming language on the Jupyter Notebook \cite{kluyver2016jupyter} using the TensorFlow libraries \cite{abadi2016tensorflow}.

As mentioned in section~\ref{sec:data_process}, each frame of an input image sequence cropped into size $h_\mathrm{in} \times w_\mathrm{in} = 250\times 250$~pixels$^2$.
With three frames, an input image sequence had the shape of $h_\mathrm{in} \times w_\mathrm{in} \times f_{\mathrm{in}} = 250 \times 250 \times 3$, where $f_{\mathrm{in}}$ is the number of frames in an input image sequence.
As for an output image sequence, each frame was cropped into size $h_\mathrm{in} \times w_\mathrm{in} = 640\times 200$~pixels$^2$.
With seven frames, an output image sequence had the shape of $h_\mathrm{out} \times w_\mathrm{out} \times f_{\mathrm{out}} = 640 \times 200 \times 7$, where $f_\mathrm{out}$ is the number of frames in an output image sequence.

In the encoder, an input image sequence is first flattened into a one-dimensional column vector: $\mathbf{X} \in \mathbb{R}^{h_\mathrm{in} \times w_\mathrm{in} \times f_{\mathrm{in}}} \to \mathbf{s}_\mathrm{in} \in \mathbb{R}^{N_\mathrm{in}}$, for $N_\mathrm{in} = 250 \times 250 \times 3 = 187,500$.
Then, it is encoded into the latent space by a linear function and a leaky rectified linear unit (ReLU) activation function.
The linear function is
\begin{equation}
\mathbf{q}_\mathrm{h} = \mathbf{W}_\mathrm{enc}\mathbf{s}_\mathrm{in} + \mathbf{b}_\mathrm{enc},
\label{eq:enc_lin}
\end{equation}
where $\mathbf{q}_\mathrm{h} \in \mathbb{R}^{N_\mathrm{enc}}$ is the output vector of the linear equation, $\mathbf{W}_\mathrm{enc} \in \mathbb{R}^{N_\mathrm{enc} \times N_\mathrm{in}}$ is the encoding weight matrix, and $\mathrm{b}_\mathrm{enc} \in \mathbb{R}^{N_\mathrm{enc}}$ is the encoding bias vector.
Here, $N_\mathrm{enc}$ is the number of encoded elements, which is set as 32 through hyperparameter tuning.
The leaky ReLU function is
\begin{equation}
s_{\mathrm{h},i} =
\cases{q_{\mathrm{h},i} & for $q_{\mathrm{h},i} \ge 0$\\
\alpha q_{\mathrm{h},i} & for $q_{\mathrm{h},i} < 0$\\}
\label{eq:enc_act}
\end{equation}
for $i = 1, \dots, N_\mathrm{enc}$, where the output of the function $s_{\mathrm{h},i}$ is an element of the encoded vector in the latent space $\mathbf{s}_{\mathrm{h}} \in \mathbb{R}^{N_\mathrm{enc}}$ and $\alpha$ is the negative-slope coefficient, which is set as $0.3$.

In the decoder, $\mathbf{s}_\mathrm{h} \in \mathbb{R}^{N_\mathrm{enc}}$ is decoded into the output layer by a linear function and a sigmoid activation function.
The linear function is
\begin{equation}
\mathbf{q}_\mathrm{out} = \mathbf{W}_\mathrm{dec}\mathbf{s}_\mathrm{h} + \mathbf{b}_\mathrm{dec},
\label{eq:dec_lin}
\end{equation}
where $\mathbf{q}_\mathrm{out} \in \mathbb{R}^{N_\mathrm{dec}}$ is the output vector of the linear equation, $\mathbf{W}_\mathrm{dec}\in \mathbb{R}^{N_\mathrm{dec} \times N_\mathrm{enc}}$ is the decoding weight matrix, and $\mathbf{b}_\mathrm{dec} \in \mathbb{R}^{N_\mathrm{dec}}$ is the decoding bias vector.
Here, $N_{\mathrm{dec}}$ is the number of decoded elements, which is the total number of pixels in a post-impact image sequence.
Thus, $N_\mathrm{dec} = h_\mathrm{out} \times w_\mathrm{out} \times f_{\mathrm{out}} = 640 \times 200 \times 7 = 896,000$.
The sigmoid function is
\begin{equation}
s_{\mathrm{out},i} = \sigma(q_{\mathrm{out},i}) = \frac{1}{1+e^{-q_{\mathrm{out},i}}},
\label{eq:dec_act}
\end{equation}
for $i = 1, \dots, N_{\mathrm{dec}}$, where the output of the function $s_{\mathrm{out},i}$ is an element of the one-dimensional column vector in the output layer $\mathbf{s}_{\mathrm{out}} \in \mathbb{R}^{N_\mathrm{dec}}$.
This function saturates negative and positive values to 0 and 1, respectively, to generate a binary number in each pixel.
Finally, the vector $\mathbf{s}_{\mathrm{out}}$ is unflattened to form the post-impact image sequences: $\mathbf{s}_\mathrm{out} \in \mathbb{R}^{N_\mathrm{dec}} \to \mathbf{Y} \in \mathbb{R}^{h_\mathrm{out} \times w_\mathrm{out} \times f_{\mathrm{out}}}$.

The prediction by the encoder--decoder can be defined as a mapping $\mathbf{Y}=F(\mathbf{X},\theta)$.
Here, $\theta$ denotes the mapping parameters that consist of all the elements of weight matrices and bias vectors: $\mathbf{W}_\mathrm{enc} \in \mathbb{R}^{N_\mathrm{enc} \times N_\mathrm{in}}$, $\mathbf{W}_\mathrm{dec} \in \mathbb{R}^{N_\mathrm{dec} \times N_\mathrm{enc}}$, $\mathbf{b}_\mathrm{enc} \in \mathbb{R}^{N_\mathrm{enc}}$, and $\mathbf{b}_\mathrm{dec} \in \mathbb{R}^{N_\mathrm{dec}}$.
Therefore, the total number of parameters $N_\mathrm{p}$ can be calculated using the following equation:
\begin{eqnarray}
N_\mathrm{p} & = N_\mathrm{enc} \times N_\mathrm{in} + N_\mathrm{dec} \times N_\mathrm{enc} + N_\mathrm{enc} + N_\mathrm{dec}\\ 
& = N_\mathrm{enc}(N_\mathrm{in} + N_\mathrm{dec} + 1) + N_\mathrm{dec},
\label{eq:N_p}
\end{eqnarray}
where $N_\mathrm{p} = 35,568,032$.

\subsection{Training}
\label{sec:training}

The training of the encoder--decoder is performed by minimizing a binary cross-entropy loss function.
The binary cross-entropy loss function was chosen because this study considers the image-generation process as a pixel-by-pixel prediction of whether a pixel is occupied by the drop morphology.
The loss $l$ was computed using the following equation:
\begin{equation}
l({\bf s}_\mathrm{true},{\bf s}_\mathrm{out})
=\sum_{i=1}^{N_\mathrm{dec}}\left[
-{s}_{\mathrm{true},i}\ln({s}_{\mathrm{out},i})-(1-{s}_{\mathrm{true},i})\ln(1-{s}_{\mathrm{out},i})\right]
\label{eq:cross_entropy},
\end{equation}
for $i = 1, \dots, N_\mathrm{dec}$, where $\mathbf{s}_\mathrm{true}$ is the one-dimensional column vectors of the actual post-impact image sequences.
When $\mathbf{s}_\mathrm{out}$ is close to $\mathbf{s}_\mathrm{true}$, $l$ will be close to 0.
On the contrary, when $\mathbf{s}_\mathrm{out}$ is not equal to $\mathbf{s}_\mathrm{true}$, $l$ increases dramatically as $\mathbf{s}_\mathrm{out}$ deviates further from $\mathbf{s}_\mathrm{true}$.

To determine how each element of the weight matrices and the bias vectors should be tweaked to minimize $l$, a backpropagation algorithm was applied to compute the gradient of the computed $l$ with respect to each element.
For the optimization, the adaptive moment (Adam) optimizer \cite{kingma2014adam} was used to effectively tweak these elements in the direction of descending gradients.

\subsection{Hyperparameter tuning}
\label{sec:hyp_tune}


Hyperparameter tuning was performed concurrently with the training of the encoder--decoder.
The goal is to find the best architecture that can achieve the desired performance with the lowest computational cost possible.

In terms of desired performance, the encoder--decoder should be able to generate the morphologies of non-splashing drops and splashing drops of different intensities.
In the case when the desired performance is not achieved, the encoder--decoder only generates the morphologies of non-splashing drops regardless of the input.

On the other hand, a low computational cost could be achieved by a simple architecture that has a small $N_p$.
For that, the number of hidden layers in the encoder and the decoder was reduced to zero.
As for $N_enc$, it was reduced to 32 by using the leaky ReLU function instead of the ReLU function for the activation of the encoded elements.
Note that the $\alpha$ of the ReLU function is 0.
The encoder--decoder could not achieve the desired performance when $N_enc$ was reduced to lower than 32.


\section{Results and discussion}
\label{sec:results}

\subsection{Interpolation prediction of trained encoder--decoder}\label{sec:gen_img_seq}


\begin{figure}[tb]
\centering
\includegraphics[width=\textwidth]{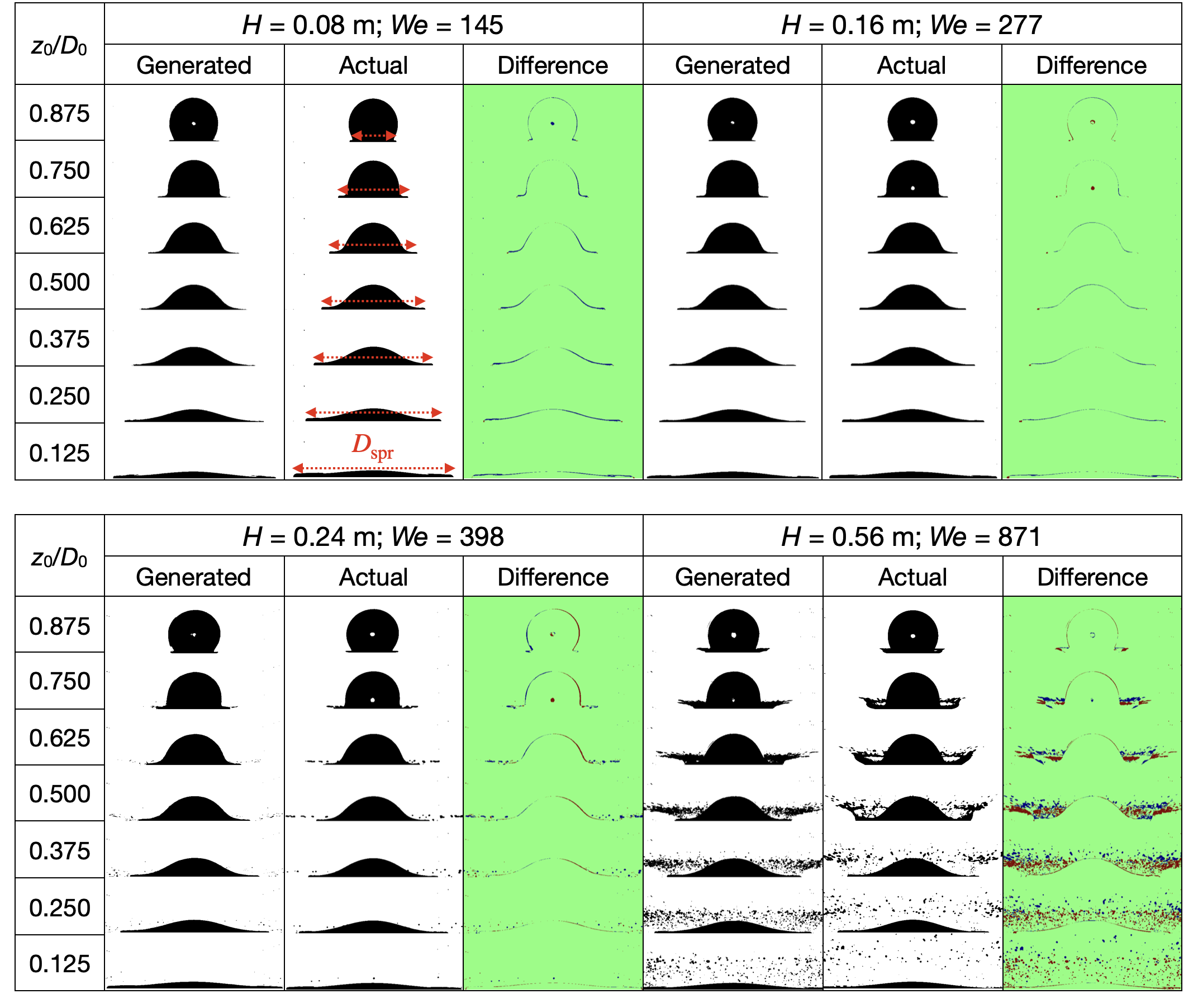}
\caption{
Examples of post-impact image sequences generated by the trained encoder--decoder and the actual binarized post-impact image sequences for the drop impact of the excluded heights: (top left) a non-splashing drop of $H = 0.08$~m and $We = 145$ (the red dotted lines in the actual image sequence show the spreading diameter $D_\mathrm{spr}$ of the impacting drop at different $z_0/D_0$); (top right) a non-splashing drop of $H = 0.16$~m and $We = 277$; (bottom left) a splashing drop of $H = 0.24$~m and $We = 398$; and (bottom right) a splashing drop of $H = 0.56$~m and $We = 871$.
The heat maps shown next to the actual image sequences show the differences between the respective generated and actual image sequences, in which green shows correctly predicted pixels, red shows pixels that are occupied by the actual drop but not by the generated drop, and blue shows the pixels that are occupied by the generated drop but not by the actual drop.
}
\label{fig:gen_act_imgSeq}
\end{figure}

Several examples of post-impact image sequences generated by the trained encoder--decoder from the pre-impact image sequences with the $H$ values reserved for testing are shown alongside the pre-impact inputs and the actual output image sequences in figure~\ref{fig:gen_act_imgSeq}.
The heat maps shown next to the actual image sequences show the differences between the respective generated and actual image sequences.
As shown, the generated and actual image sequences are generally in good agreement.
For non-splashing drops in particular, the difference is represented only by a thin line along the contour of the drop.

For splashing drops of $H = 0.56$~m, careful examination of the generated and actual image sequences reveals a difference in the ejection angle of the secondary droplets.
This can be seen from the example shown in figure~\ref{fig:gen_act_imgSeq} (bottom right) in the splashing drop of $H = 0.56$~m and $We = 871$.
Especially for $0.500 \leq z_0/D_0 \leq 0.750$ (the second to fourth frames), the secondary droplets in the generated image sequences are ejected outwards, forming a shape like a plate, while those in the actual image sequences are ejected upwards, forming a shape like a bowl.
Nevertheless, the results are satisfying because there is a clear difference in the distribution and number of ejected secondary droplets between the image sequences generated using different $H$ values.
This can be seen when comparing the splashing drops of $H = 0.24$~m and $We = 398$ (bottom left) and $H = 0.56$~m and $We = 871$ (bottom right).
These findings demonstrate the ability of the trained encoder--decoder to generate image sequences of splashing drops of different morphologies resulting from different physical parameters.

\subsection{Spreading diameter of impacting drop}\label{sec:spr_rad}

For quantitative evaluation of the generated post-impact image sequences, the spreading diameter $D_\mathrm{spr}$ of the impacting drop in each frame ($z_0/D_0$) of the image sequence generated by the trained encoder--decoder for each set of testing data was measured according to the definition shown in figure~\ref{fig:gen_act_imgSeq} and normalized by $D_0$.

\begin{figure}[tb]
\centering
\begin{subfigure}{0.45\textwidth}
\includegraphics[width=\textwidth]{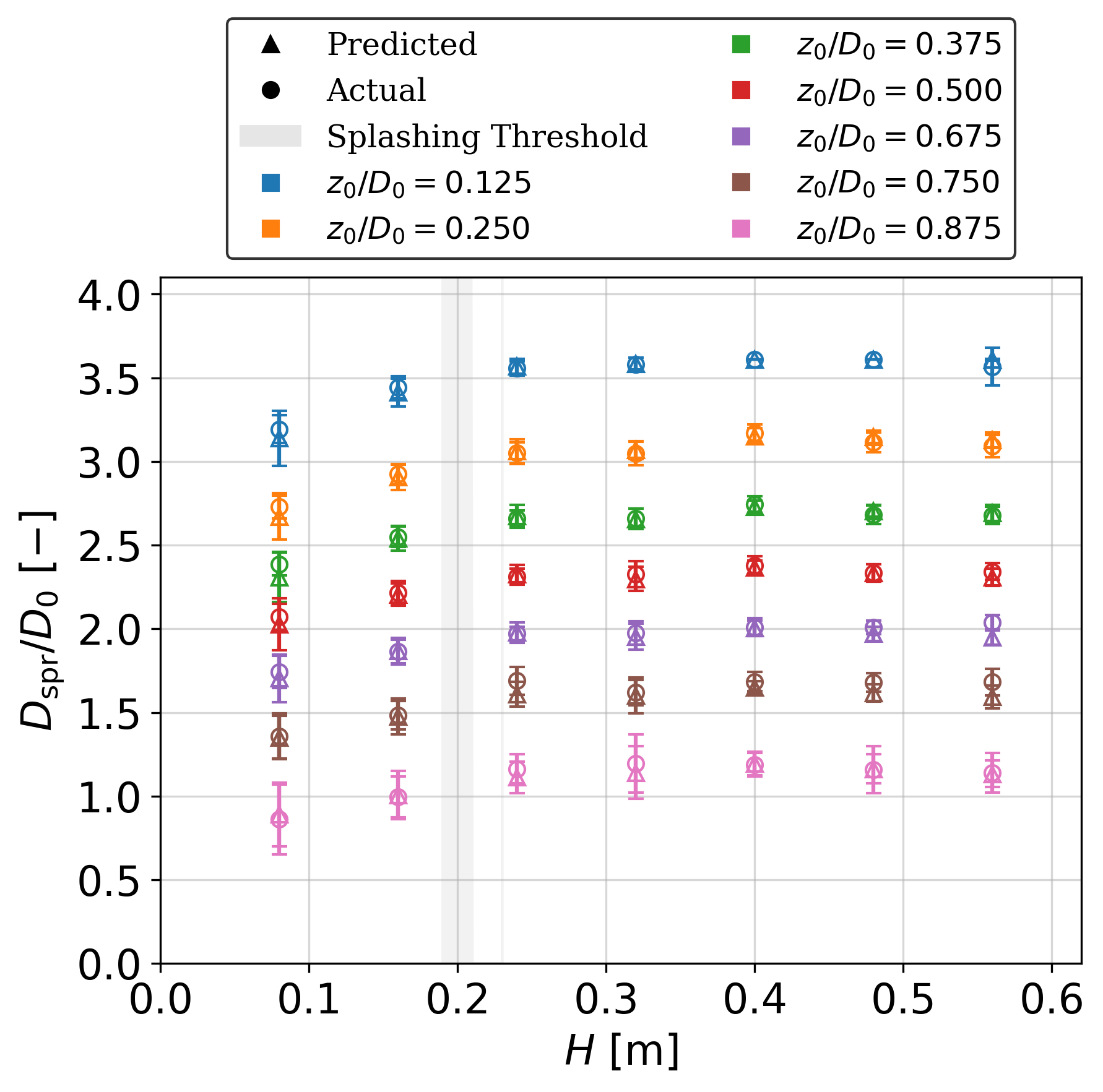}
\caption{}\label{fig:sprD_H}
\end{subfigure}
\hfill
\centering
\begin{subfigure}{0.45\textwidth}
\includegraphics[width=\textwidth]{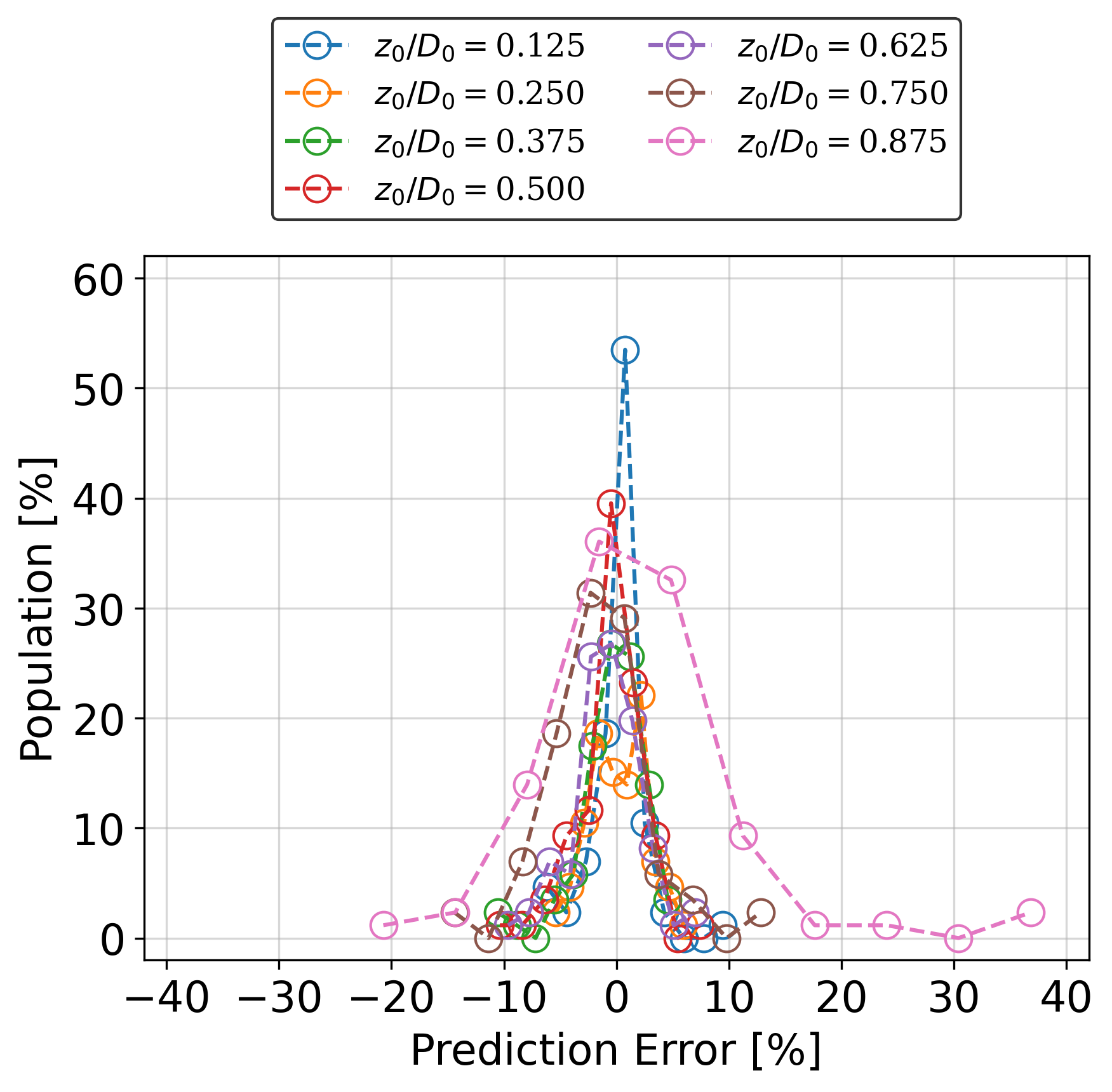}
\caption{}\label{fig:sprD_hist_z0}
\end{subfigure}
\\
\centering
\begin{subfigure}{0.45\textwidth}
\includegraphics[width=\textwidth]{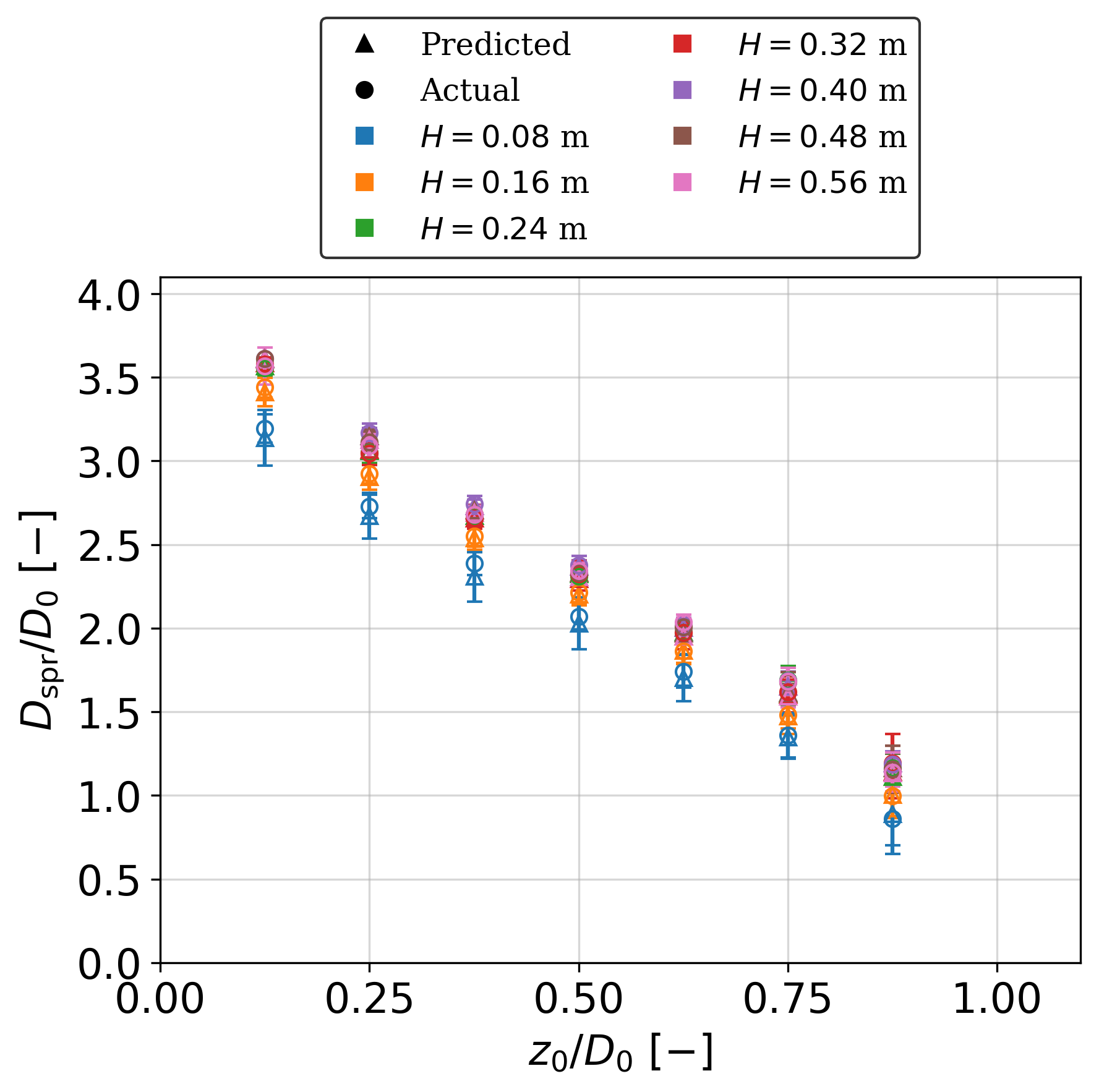}
\caption{}\label{fig:sprD_z0}
\end{subfigure}
\hfill
\begin{subfigure}{0.45\textwidth}
\includegraphics[width=\textwidth]{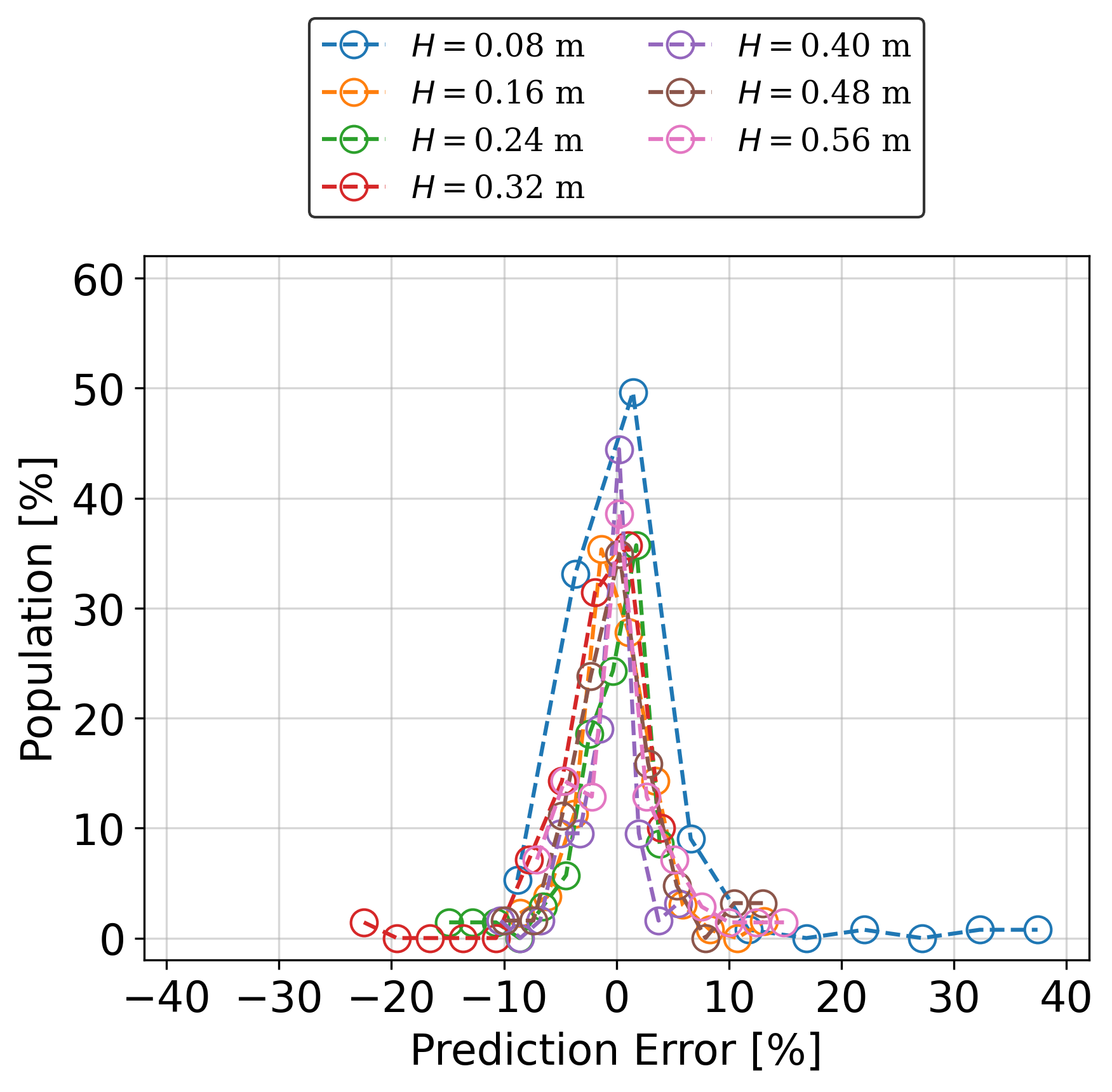}
\caption{}\label{fig:sprD_hist_H}
\end{subfigure}
\caption{
(\protect\subref*{fig:sprD_H})~Mean normalized spreading diameters $D_\mathrm{spr}/D_0$ of the impacting drops averaged among each frame ($z_0/D_0$) of the generated and actual image sequences for interpolation impact heights $H$ and
(\protect\subref*{fig:sprD_hist_z0})~plots of the prediction errors.
(\protect\subref*{fig:sprD_z0})~Mean $D_\mathrm{spr}/D_0$ values of the generated and actual image sequences averaged for each interpolation impact height $H$ and (\protect\subref*{fig:sprD_hist_H})~plots of the prediction errors.
}
\label{fig:sprD}
\end{figure}

Figure~\ref{fig:sprD_H} shows the mean normalized spreading diameters $D_\mathrm{spr}/D_0$ of the impacting drops averaged among each frame of the generated and actual image sequences of the interpolation impact heights $H$, with the error bars showing the standard deviations.
In figure~\ref{fig:sprD_H}, for each interpolation $H$ value, the mean $D_\mathrm{spr}/D_0$ values of the generated and actual image sequences of the impacting drop for each $z_0/D_0$ overlap, indicating excellent agreement between the two.
The prediction error is quantified by computing the percentage difference between the $D_{\mathrm{spr}}$ values of the impacting drop in each frame of the generated and actual image sequences.

Plots of the prediction error for each $z_0/D_0$ are shown in figure~\ref{fig:sprD_hist_z0}.
Other than for $z_0/D_0 = 0.875$, all of the prediction errors are low: between $-15$\% and $15$\%.
The prediction error for $z_0/D_0 = 0.875$ is relatively high because the lamella is not fully developed; for some non-splashing drops, a lamella is not ejected, whereas for some splashing drops, it is lifted up.
These phenomena were considered by Riboux and Gordillo \cite{riboux2017boundary,riboux2014experiments}, who stated that the ejection time of the lamella scales with $We$ when the Ohnesorge number $Oh$ is sufficiently small, and they attributed splashing to the lift force acting on the lamella.

To further analyse the behaviour of $D_\mathrm{spr}$, the $D_\mathrm{spr}/D_0$ values of the generated and actual image sequences were averaged among each interpolation $H$ value and plotted against $z_0/D_0$, as shown in figure~\ref{fig:sprD_z0}, in which the error bars show the standard deviations.
The distribution shows that there is a negative correlation between $D_\mathrm{spr}/D_0$ and $z_0/D_0$.
This is because as time elapses, the central height of the drop $z_0$ decreases as the lamella continues to spread, thus $D_\mathrm{spr}$ increases.
Similar to figure~\ref{fig:sprD_H}, for all values of $z_0/D_0$, the mean $D_\mathrm{spr}/D_0$ values of the generated and actual image sequences of the impacting drop for each interpolation $H$ value overlap, indicating excellent agreement between the two.

Plots of the prediction error for each $H$ value are shown in figure~\ref{fig:sprD_hist_H}.
Other than for $H = 0.08$ and $0.32$~m, all of the prediction errors are again low: between $-20$\% and $20$\%.
The prediction error of $H = 0.08$~m is relatively high because the lamella is ejected later due to the low $We$ value.
Thus, the standard deviation of $D_\mathrm{spr}$ is higher, especially for $z_0/D_0 = 0.875$.

\subsection{Splashing/non-splashing prediction}\label{sec:classification}
The accuracy of splashing/non-splashing prediction of the trained encoder--decoder was evaluated by inspecting the impacting drop in each frame ($z_0/D_0$) of the generated image sequences.
To determine whether an impacting drop is splashing, the main criteria are the presence of ejected secondary droplets and the behaviour of the lamella.
Although a splashing drop is mainly characterized by the ejection of secondary droplets, for some splashing drops, there are very few detached secondary droplets at the early stage of the impact when $z_0/D_0 = 0.75$ (see bottom left of figure~\ref{fig:gen_act_imgSeq}).
Nevertheless, the splashing of these drops can still be identified by the lifted lamella \cite{riboux2014experiments}.

\begin{figure}[tb]
\centering
\includegraphics[width=0.6\textwidth]{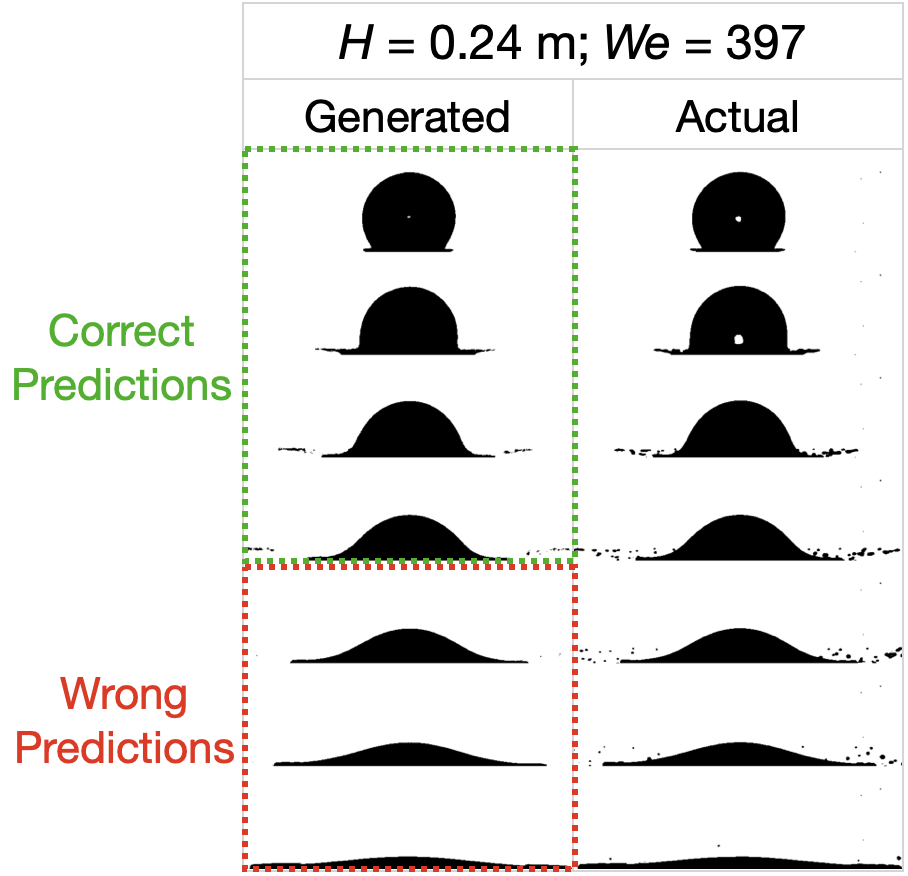}
\caption{Post-impact image sequences generated by the trained encoder--decoder and actual binarized post-impact image sequence of a splashing drop with $H = 0.24$~m and $We = 397$.
The encoder--decoder correctly generated images showing the morphologies of splashing drops for $z_0/D_0 = 0.875$, $0.750$, $0.625$, and $0.500$, but incorrect morphologies of non-splashing drops were generated for $z_0/D_0 = 0.375$, $0.250$, and $0.125$.
}
\label{fig:diff_pred}
\end{figure}

It is important to point out that the splashing/non-splashing prediction is not necessarily the same for all of the frames in a generated image sequence.
One example is the image sequence generated for the splashing drop of $H = 0.24$~m and $We = 397$ that is shown alongside the actual image sequence in figure~\ref{fig:diff_pred}.
In this figure, the encoder--decoder correctly generated images that show the morphologies of splashing drops for $z_0/D_0 = 0.875$, $0.750$, $0.625$, and $0.500$, but it generated images showing incorrect morphologies of non-splashing drops for $z_0/D_0 = 0.375$, $0.250$, and $0.125$.
Thus, instead of evaluating the splashing/non-splashing prediction of a generated image sequence as a whole, it is necessary to perform a frame-by-frame or $z_0/D_0$-by-$z_0/D_0$ evaluation.

\begin{figure}[tb]
\centering
\begin{subfigure}{0.58\textwidth}
\includegraphics[width=\textwidth]{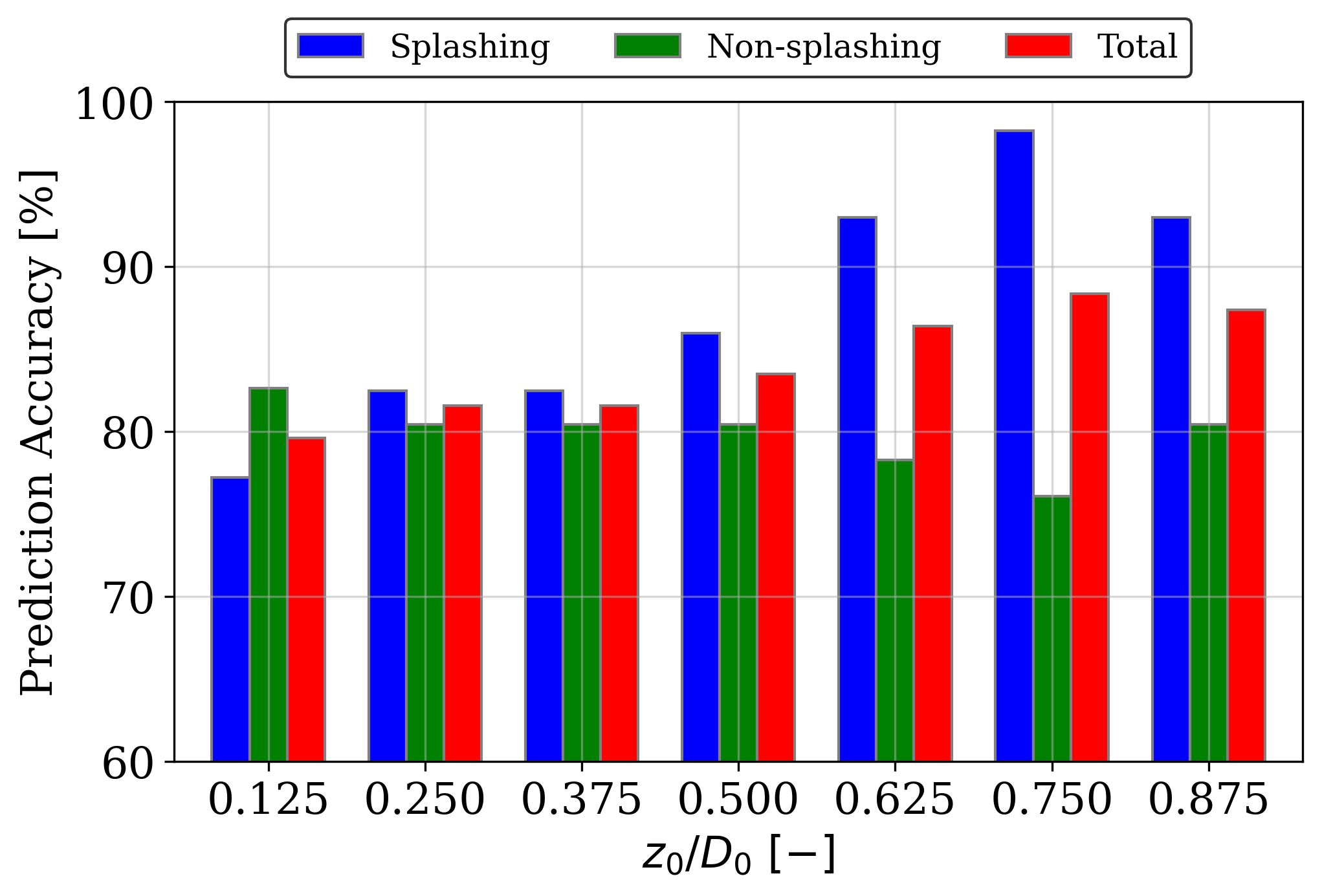}
\caption{}\label{fig:predAcc_z0}
\end{subfigure}
\\
\begin{subfigure}{0.58\textwidth}
\includegraphics[width=\textwidth]{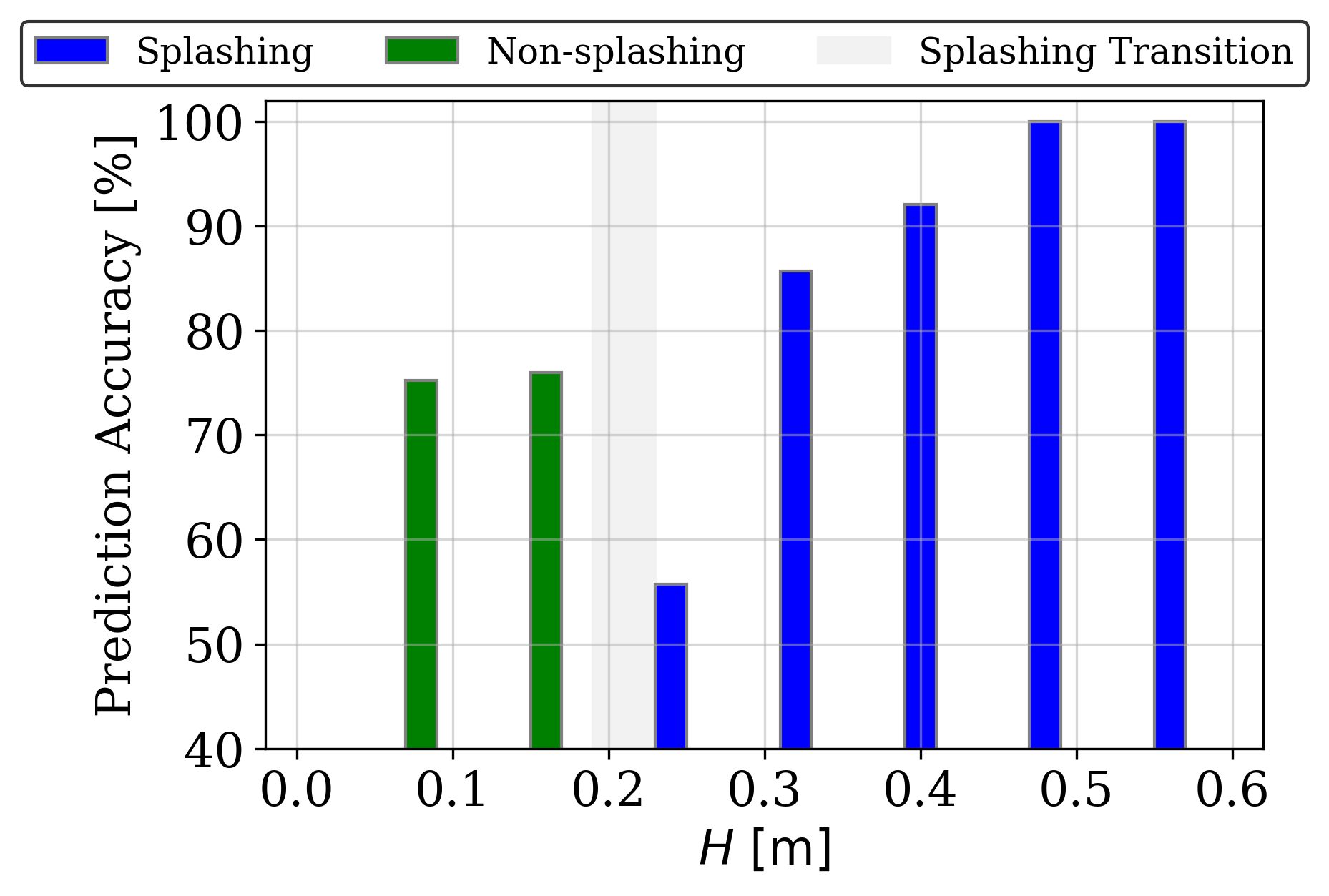}
\caption{}\label{fig:predAcc_H}
\end{subfigure}
\\
\begin{subfigure}{0.58\textwidth}
\includegraphics[width=\textwidth]{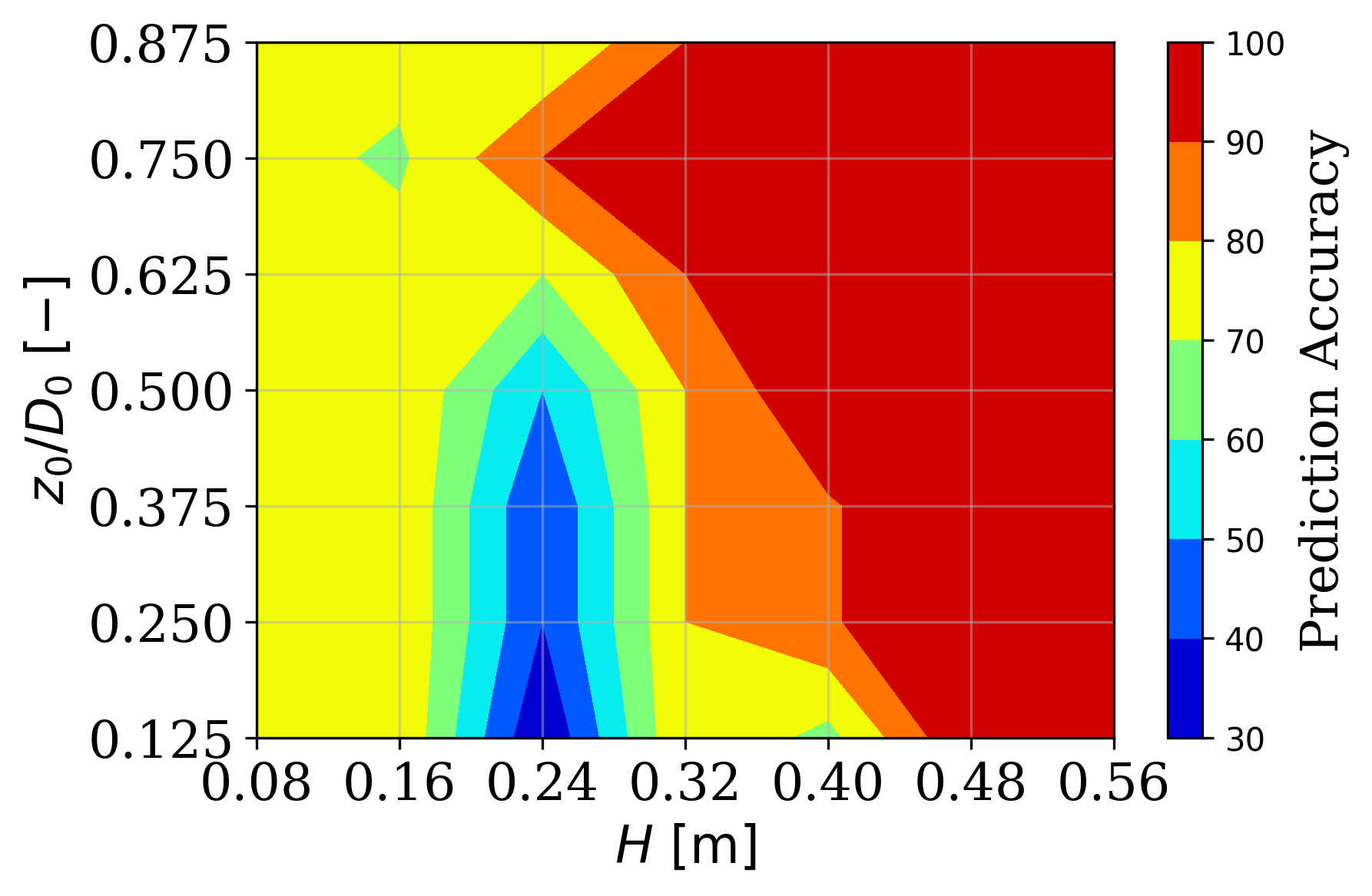}
\caption{}\label{fig:acc_cb}
\end{subfigure}
\caption{
Splashing/non-splashing prediction accuracy of the trained encoder--decoder:
(\protect\subref*{fig:predAcc_z0})~for each frame ($z_0/D_0$) of the generated image sequences; 
(\protect\subref*{fig:predAcc_H})~for each interpolation impact height $H$; and
(\protect\subref*{fig:acc_cb})~for each $z_0/D_0$ at each interpolation $H$ value.
}
\label{fig:predAcc}
\end{figure}

The accuracy of splashing/non-splashing prediction for each value of $z_0/D_0$ of the generated image sequences is shown by the bar chart in figure~\ref{fig:predAcc_z0}.
A high prediction accuracy was achieved, with an overall accuracy level greater than 80\%.
For all $z_0/D_0$ values except $0.125$, the splashing prediction accuracy was higher than the non-splashing prediction accuracy.
This is because when $z_0/D_0 = 0.125$, the ejected secondary droplets are widely scattered away from the impacting drop, and it is therefore difficult for the trained encoder--decoder to capture and reproduce this.
However, when $z_0/D_0 \ge 0.625$, the secondary droplets are only just ejected from the impacting drop and are still accumulating around the lamella; thus, the splashing prediction accuracies are higher than 90\%.

The accuracy of splashing/non-splashing prediction for each interpolation impact height $H$ was also evaluated, and the results are shown in the bar chart in figure~\ref{fig:predAcc_H}.
Note that the accuracy was computed frame by frame for every generated image sequence.
For example, for the generated image sequences shown in figure~\ref{fig:diff_pred}, four out of the seven frames were correctly predicted, hence the accuracy was 57\%.
Since there is a splashing transition within $0.20~\mathrm{m} \leq H \leq 0.22$~m, there are only non-splashing and splashing prediction accuracies for $H \leq 0.16$~m and $H \ge 0.24$~m, respectively.
Similar to figure~\ref{fig:predAcc_z0}, the splashing prediction accuracy is higher than the non-splashing prediction accuracy, except for at $H = 0.24$~m.
The prediction accuracy is very low for $H = 0.24$~m because this is the lowest of the splashing $H$ values.
In other words, the impact velocity is the lowest for $H = 0.24$~m, and this means that splashing features such as the number of ejected secondary droplets are least prevalent.
Nevertheless, the splashing prediction accuracy increases with $H$ and eventually reaches a perfect $100$\% for $H \ge 0.48$~m.

A more detailed analysis was performed, and the prediction accuracy for each value of $z_0/D_0$ at each interpolation $H$ value is shown by the heat map in figure~\ref{fig:acc_cb}.
Similar to Figs.~\ref{fig:predAcc_z0} and \ref{fig:predAcc_H}, it can be seen that the splashing prediction accuracy ($H \ge 0.24$~m) is higher than the non-splashing prediction accuracy ($H \leq 0.16$~m), except for $H = 0.24$~m, which has the lowest prediction accuracy.
Additionally, the prediction accuracy is a perfect 100\% for each $z_0/D_0$ for interpolation $H \ge 0.48$~m.
For splashing $H$ values, the prediction accuracy tends to decrease with $z_0/D_0$, while for non-splashing $H$ values, the prediction accuracy tends to increase with $z_0/D_0$.
This shows that the trained encoder--decoder tends to predict a splashing drop for high $z_0/D_0$ values and a non-splashing drop for low $z_0/D_0$ values.

\subsection{Analysis on the prediction process}
\label{sec:pred_process}

Analysis was performed to understand the prediction process by the encoder--decoder, specifically, the image features of the pre-impact image sequences that the trained encoder identifies and how these image features affect the generation of the post-impact image sequences by the trained decoder.
However, there are too many parameters involved in the prediction process, the important parameters had to be first identified by extracting the important encoded elements.

\begin{figure}
\centering
\includegraphics[width=\columnwidth]{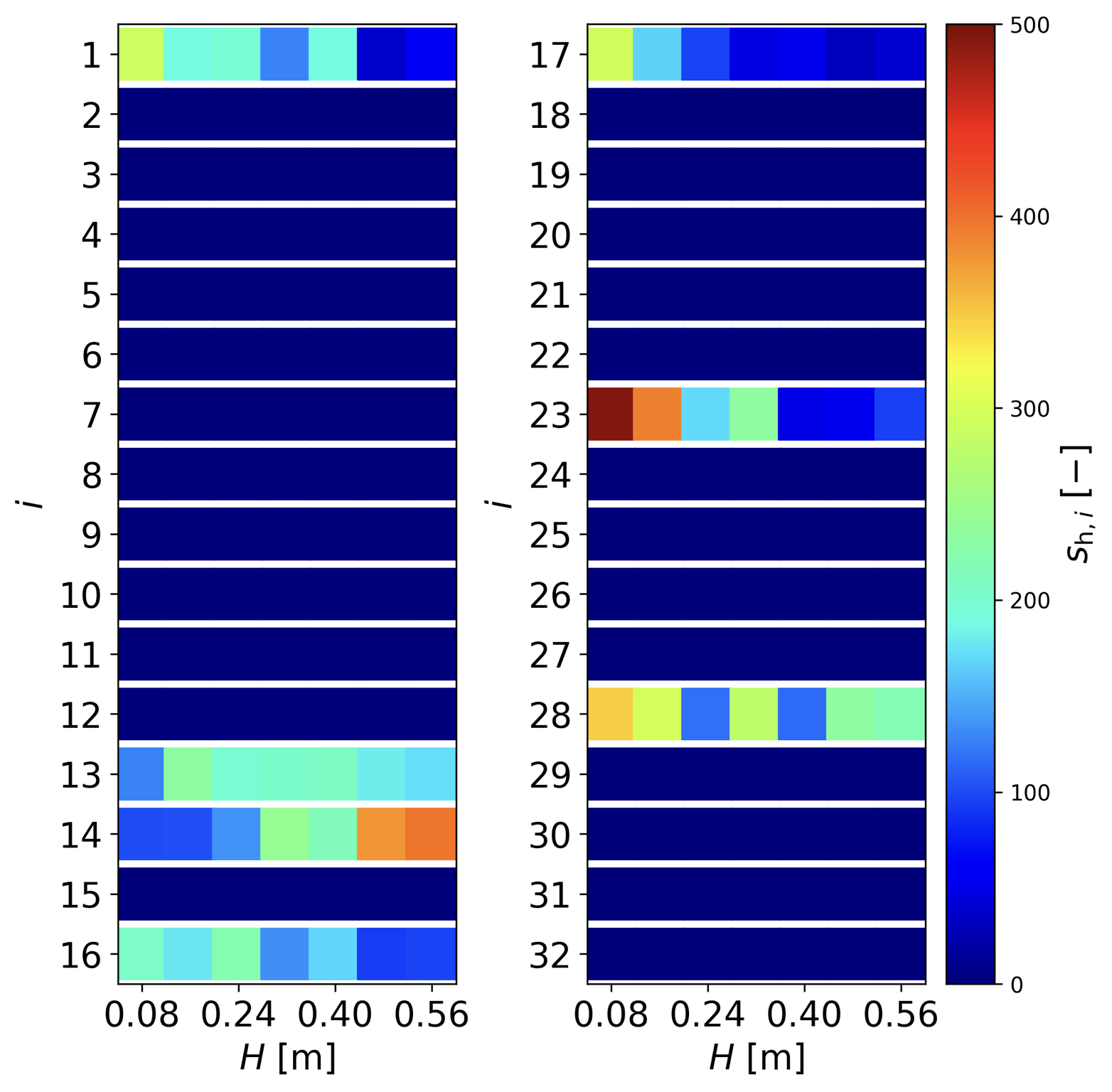}
\caption{
The value of each encoded element computed for the test image sequences averaged among splashing and non-splashing drops.
}
\label{fig:ele_imp}
\end{figure}

The encoded elements, which are shown by the green nodes in figure~\ref{fig:enc_dec}, play important roles as they connect the encoder and the decoder.
The important elements were extracted by analyzing the $s_{\mathrm{h},i}$ computed for the test image sequences.
Here, $i = 1, \dots, N_\mathrm{enc}$, where $N_\mathrm{enc}$ was set to 32.
Based on the pre-impact image sequences, the encoder computes the values of $s_{\mathrm{h},i}$.
The values of $s_{\mathrm{h},i}$ were averaged among the test image sequences of the same $H$ and plotted in figure~\ref{fig:ele_imp}.
The elements with the value $s_{\mathrm{h}} \approx 0$ were inactive as they did not affect the computation of $s_{\rm{out},i}$, i.e. the values of the intensity $s_{\rm{out},i}$ in the post-impact image sequence that show the drop morphology.
On the other hand, the active elements are $i =$ 1, 13, 14, 16, 17, 23, and 28.
These active elements are important because their values were used by the decoder to compute the values of $s_{\rm{out},i}$.

Instead of all the encoded elements, the explanation of the prediction process of the trained encoder-decoder here focuses on the analysis of the elements of $i =$ 14 and 23.
These two elements are good examples because they have the highest range of $s_{\mathrm{h},i}$.
Besides, their relationships with $H$ are opposite to one another.
As $H$ increases from 0.08 to 0.56~m, $s_{\mathrm{h},14}$ increases from $\approx 100$ to $\approx 500$ while $s_{\mathrm{h},23}$ decreases from $\approx 500$ to $\approx 100$.
These indicate that a splashing drop has a high value of $s_{\mathrm{h},14}$ and a low value of $s_{\mathrm{h},23}$ while a non-splashing drop has a low value of $s_{\mathrm{h},14}$ and a high value of $s_{\mathrm{h},23}$.

\begin{figure}
\centering
\includegraphics[width=\columnwidth]{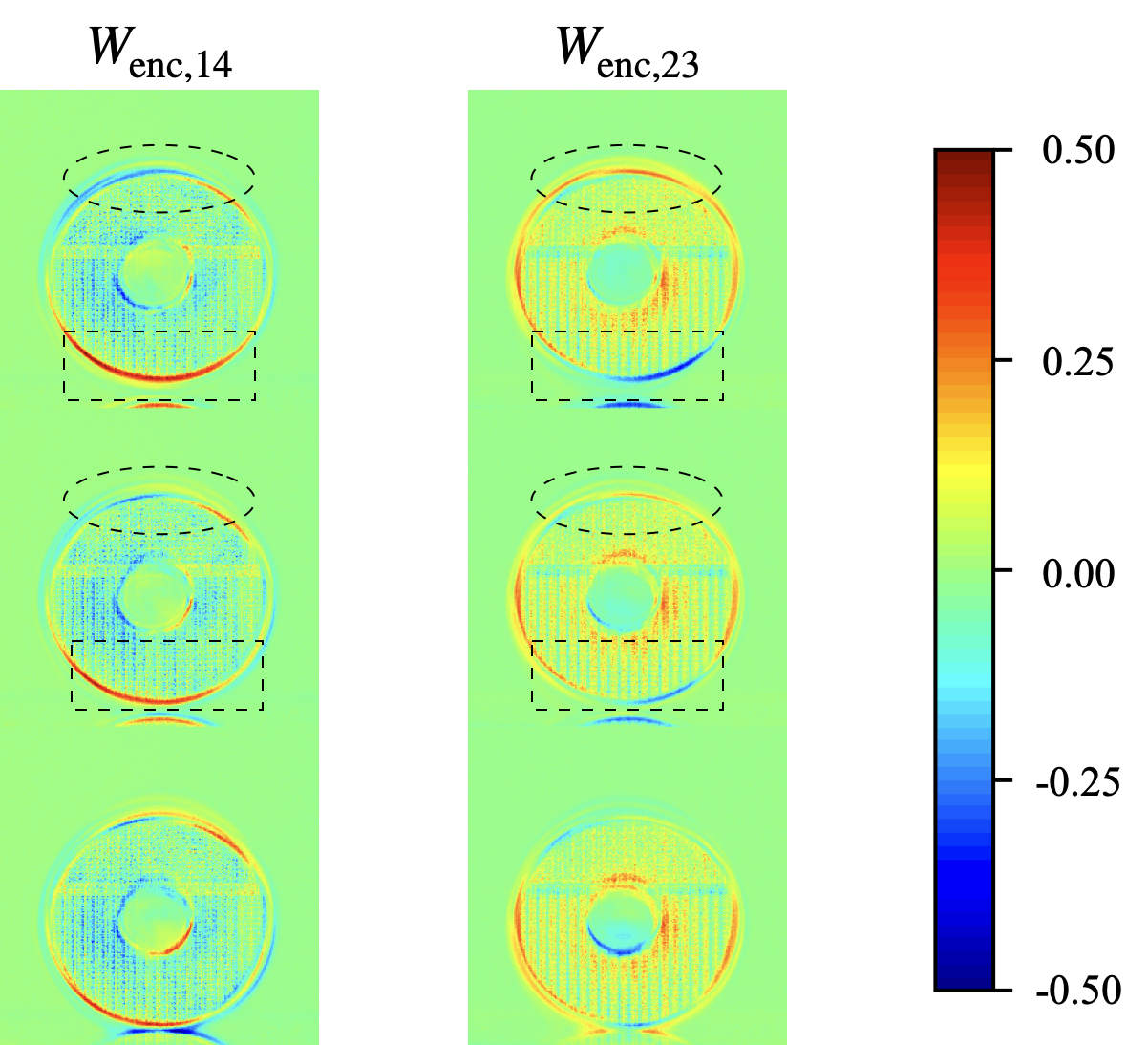}
\caption{
Colourmaps of the reshaped encoding weight elements of $i =$ 14 and 23. In the first two frames, the areas in the dashed circles correspond to those covered by splashing drops but not non-splashing drops, while the areas in the dashed rectangles correspond to those not covered by splashing drops but non-splashing drops.
}
\label{fig:w_enc}
\end{figure}

To understand the image features extracted by the encoder from the pre-impact image sequences, the elements of the encoding weight matrix $\mathbf{W}_\mathrm{enc}$ of $i =$ 14 and 23 were visualized.
For this purpose, the encoding weight elements $w_{\mathrm{enc},14}$ and $w_{\mathrm{enc},23}$ were reshaped into the shape of a pre-impact image sequence and presented in figure~\ref{fig:w_enc} with the blue-green-red (BGR) scale from -0.50 to 0.50.
Since splashing drops have a higher $U_0$ than non-splashing drops, the splashing drops cover higher positions than the non-splashing drops in the first two frames of a pre-impact image sequence.
The respective positions are shown in figure~\ref{fig:w_enc}, where the dashed circles show the area covered by splashing drops but not by non-splashing drops, while the dashed rectangles show the area not covered by splashing drops but by non-splashing drops.
Observation on these areas shows that in the colourmap of $w_{\mathrm{enc},14}$, there are more negative values (blue) in the dashed circles, while there are more positive values (red) in the dashed rectangles.
Thus, the negative values remain for a non-splashing drop leading to a lower value of $s_{\mathrm{h},14}$, while the positive values remain for a splashing drop leading to a higher value of $s_{\mathrm{h},14}$.
On the other hand, in the colourmap of $w_{\mathrm{enc},23}$, there are more positive values (red) in the dashed circles, while there are more negative values (blue) in the dashed rectangles.
Thus, the negative values remain for a non-splashing drop leading to a lower value of $s_{\mathrm{h},23}$, while the positive values remain for a splashing drop leading to a higher value of $s_{\mathrm{h},23}$.

\begin{figure}
\centering
\includegraphics[width=\columnwidth]{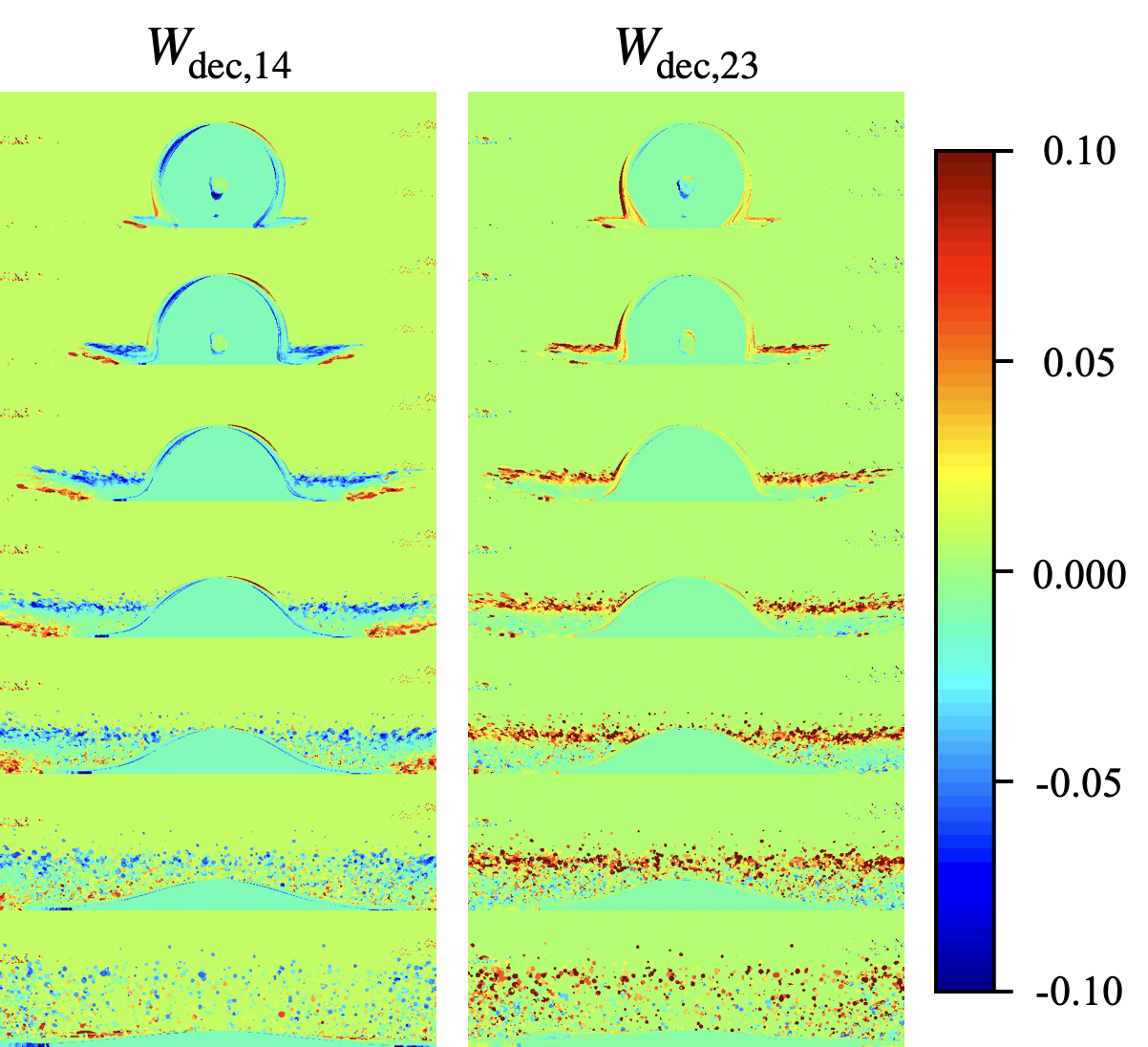}
\caption{
Colourmaps of the reshaped decoding weight elements of $i =$ 14 and 23.
}
\label{fig:w_dec}
\end{figure}

To understand how the decoder computes the intensity values of each pixel in the post-impact image sequences to predict the drop morphology, the elements of the decoding weight matrix $\mathbf{W}_\mathrm{dec}$ of $i =$ 14 and 23 were visualized.
For this purpose, the decoding weight elements $w_{\mathrm{dec},14}$ and $w_{\mathrm{dec},23}$ were reshaped into the shape of a post-impact image sequence and presented as colourmaps in figure~\ref{fig:w_dec} with the blue-green-red (BGR) scale from -0.10 to 0.10.
Analysis was then performed on $w_{\mathrm{dec},14}$ and $w_{\mathrm{dec},23}$.

Since the value of $s_{\mathrm{h},14}$ is high for splashing drops but low for non-splashing drops, there are more negative values (blue) in the colourmaps of $w_{\mathrm{dec},14}$.
For splashing drops, since $s_{\mathrm{h},14}$ is high, these negative values remain and become zero after being activated by the sigmoid function (Eq.~\ref{eq:dec_act}).
These zero values form the morphology of a drop.
On the other hand, for non-splashing drops, since $s_{\mathrm{h},14}$ is low, these negative values do not remain but become non-zero values after being activated by the sigmoid function.
These non-zero values do not form the morphology of a drop.
Note that there are some positive values under the distribution of negative values that forms the morphology of a lamella.
This can be related to the higher ejection angle of the lamella of a corona splash, which occurs at a higher velocity than a prompt splash.

A similar but opposite explanation can be made for the element of $i =$ 23.
Since the value of $s_{\mathrm{h},23}$ is low for splashing drops but high for non-splashing drops, there are more positive values (red) in the colourmaps of $w_{\mathrm{dec},23}$.
For splashing drops, since $s_{\mathrm{h},23}$ is low, these positive values do not remain but become zero values that form the morphology of a drop.
On the other hand, for non-splashing drops, since $s_{\mathrm{h},14}$ is high, these positive values remain and become non-zero values, thus not forming the morphology of a drop.



\section{Conclusion}\label{sec:conclusion}

In this study, an encoder--decoder model was trained to generate videos in the form of image sequences that can accurately represent the actual morphological evolutions of splashing and non-splashing drops during their impact on a solid surface under different impact velocities.

Interpolation prediction was performed by reserving the image sequences of drops of certain interpolation impact heights $H$ for testing.
The image sequences generated for the interpolation $H$ values showed good agreement with the actual image sequences.
Notably, among splashing drops of different $H$, there was a clear difference in morphology in terms of the distribution and number of ejected secondary droplets.

The spreading diameters $D_\mathrm{spr}$ of the impacting drops in the generated image sequences showed excellent agreement with those in the actual image sequences.
Averaging among the frames of each normalized central height of the impacting drops $z_0/D_0$, the prediction error was $\pm 15\%$ for $z_0/D_0 \leq 0.750$.
Averaging among $H$, the prediction error was $\pm 20\%$ for $H$ values other than $0.08$ and $0.32$~m.
The prediction error can be related to the development of the lamella, and the error is higher at the beginning of the lamella ejection.

The overall accuracy of splashing/non-splashing prediction of the generated image sequences was greater than 80\%.
The trained encoder--decoder tends to generate images of splashing drops for frames of higher $z_0/D_0$, while for frames of lower $z_0/D_0$, it tends to generate images of non-splashing drops.
This is because more ejected secondary droplets accumulate around an impacting splashing drop when $z_0/D_0$ is high; thus, those splashing features can be easily captured by the trained encoder--decoder.
Conversely, when $z_0/D_0$ is low, the ejected droplets are widely scattered away from the impacting drop, and it is thus difficult for this to be captured by the trained encoder--decoder.

The findings of this study demonstrate the ability of the trained encoder--decoder to generate image sequences that can accurately represent the morphologies of splashing and non-splashing drops for intermediate data points.
This approach provides a faster and cheaper alternative to experiments and numerical simulations.
The ability of the trained encoder--decoder could be developed for researching other phenomena, especially multiphase flows.

\section*{References}
\bibliographystyle{iopart-num}
\bibliography{iopart-num}

\end{document}